\RequirePackage[2020-02-02]{latexrelease}
\documentclass[11pt,prd,aps,showpacs,eqsecnum,floatfix,nofootinbib,preprint,tightenlines]{revtex4}
\usepackage{graphicx}				
\usepackage{latexsym}
\usepackage{epsf}
\usepackage{graphicx}
\usepackage{slashed}
\usepackage{amsmath}
\usepackage{amssymb}
\usepackage{color}
\usepackage{colordvi}
\usepackage{multirow}
\newcommand{\be}{\begin{equation}}
\newcommand{\ee}{\end{equation}}
\newcommand{\bea}{\begin{eqnarray}}
\newcommand{\eea}{\end{eqnarray}}
\newcommand{\ba}{\begin{array}}
\newcommand{\ea}{\end{array}}
\newcommand{\nn}{\nonumber}

\newcommand{\tr}{{\rm tr}}
\newcommand{\rmd}{{\rm d}}
\newcommand{\rmO}{{\rm O}}
\newcommand{\Ep}{E_{\pi,\vec{p}}}

\newcommand{\gpi}{g_{\pi}}
\newcommand{\gmid}{g^{\rm mid}_{\pi}(t)}
\newcommand{\gsum}{g^{\rm sum}_{\pi}(t)}

\newcommand{\tba}{\tilde{\beta}_1}
\newcommand{\ntba}{{\beta}_{1}}

\newcommand{\pref}[1]{(\ref{#1})}

\newcommand{\talo}[1]{\tilde{\alpha}^{(0)}_{#1}}
\newcommand{\anlo}[1]{\alpha^{(1)}_{#1}}
\newcommand{\DMeff}{\Delta M_B^{B\pi}}
\newcommand{\fhat}{\hat{f}}
\newcommand{\fhatsqr}{\hat{f}^2}
\newcommand{\fhatsqreff}{\fhatsqr_{\rm eff}}
\newcommand{\fhateff}{\fhat^{LS}_{\rm eff}}

\newcommand{\Dfhateff}{\Delta\fhat^{B\pi}}
\newcommand{\Dgsum}{\Delta g^{{\rm sum},B\pi}_{\pi}}
\newcommand{\Dgmid}{\Delta g^{{\rm mid},B\pi}_{\pi}}

\newcommand{\msbar}{\overline{\mathrm{MS}}}
\newcommand{\rowheight}{\rule[-3mm]{0mm}{8mm}}

\begin{document}

\renewcommand{\thefootnote}{$*$}

\preprint{HU-EP-23/12-RTG}

\title{$B\pi$ excited-state contamination in lattice calculations of {\em B}-meson correlation functions}

\author{Oliver B\"ar$^{a}$, Alexander Broll$^{a}$ and Rainer Sommer$^{a,b}$} 
\affiliation{$^a$Institut f\"ur Physik,
\\Humboldt Universit\"at zu Berlin,
\\12489 Berlin, Germany\\
$^b$NIC, DESY\\
Platanenallee 6,
\\15738 Zeuthen, Germany\\
}

\begin{abstract}
Multi-particle states with additional pions are expected to result in a difficult-to-control excited-state contamination in lattice simulations. We show that heavy meson chiral perturbation theory can be employed to estimate the contamination due to two-particle $B\pi$ states in various $B$-meson observables like the $B$-meson decay constant and the $BB^*\pi$ coupling. We work in the static limit and to next-to-leading order in the chiral expansion, i.e. including $\rmO(p)$. The $B\pi$ states are found to typically overestimate the observables at the few percent level. We determine two of the LECs from $B\to \pi$ form factor computations of the KEK group \cite{Colquhoun:2022atw} and discuss ways to determine the others.
In particular two LECs which are associated with smeared interpolating fields 
seem to be  easily accessible and thus open up a way to systematically study the effect of
smearing on  excited state effects.
\end{abstract}

\maketitle

\renewcommand{\thefootnote}{\arabic{footnote}} \setcounter{footnote}{0}

\newpage
%
\section{Introduction} 
%
Decays and other properties of $B$-mesons have long 
been studied because they allow access to otherwise elusive parameters of the Standard Model \cite{Belle-II:2018jsg}. In particular they allow for consistency tests of the Standard Model, such as the unitarity of the CKM matrix. 
Many $B$-decays have rather small decay rates in the Standard Model, so they often compete with
non-standard physics and there is a potential to discover the latter.
Analysing experimental $B$-physics results in terms
of the Standard Model requires non-perturbative control of the theory by lattice QCD~\cite{Boyle:2022uba,Kanekolatt2023}. There are a number of challenges to compute $B$-physics observables from lattice QCD with sufficient precision. Essentially they can be traced back to the large ratio between the relevant physics scales $m_B \approx 5$~GeV and $m_\pi\approx 140$~MeV but also to the need to control excited-state contaminations~\cite{Hashimoto:2018umc,Bahr:2019eom,Egerer:2020hnc}.

In a series of papers \cite{Bar:2015zwa,Bar:2016uoj,Bar:2016jof,Bar:2018wco,Bar:2018xyi,Bar:2019gfx,Bar:2019igf,Bar:2021crj}, chiral perturbation theory (ChPT) \cite{Weinberg:1978kz,Gasser:1983yg,Gasser:1984gg}, the low-energy effective theory of QCD, was employed to study the  excited-state contamination due to 2-particle nucleon-pion ($N\pi$) states in lattice estimators for a variety of nucleon observables. 
In many cases the results were found to describe surprisingly well  discrepancies between the lattice estimates and the phenomenologically expected results. The impact of $N\pi$ states is particularly large in the axial and pseudoscalar form factors of the nucleon. In that case  
ChPT provides an understanding for the so-called PCAC puzzle \cite{Rajan:2017lxk,Bali:2018qus}: the apparent violation of the generalised Goldberger-Treiman relation between the axial and pseudoscalar form factors is caused by the contribution of a low-energetic $N\pi$ state in the induced pseudoscalar form factor \cite{Bar:2018xyi}. ChPT also showed  \cite{Bar:2019igf}  that the projection method proposed in Ref.\ \cite{Bali:2018qus} is insufficient to solve the PCAC puzzle, and it guided the development of alternative strategies \cite{Jang:2019vkm,Bali:2019yiy} to deal with the $N\pi$ contamination.

In this paper we extend these investigations  to $B$-meson observables. Employing Heavy Meson (HM) ChPT \cite{Wise:1992hn,Burdman:1992gh} we study the $B\pi$ excited-state contamination in estimators for  the $B$-meson mass, the decay constant and the $B^*B\pi$ coupling. The basic idea and principles are the same as in the nucleon sector, but the particle content and symmetry properties of the chiral effective theory are different. 

We consider QCD with a heavy $b$-quark in the static approximation and perform the SU(2) chiral expansion; the strange quark is considered heavy. In contrast to the nucleon example we here work through next-to-leading order (NLO) in this expansion. This means that corrections linear in spatial
momenta and energies of the pions are taken into account, while $E_\pi^2,m_\pi^2$ terms are neglected.
At LO the results depend on one low-energy coefficient (LEC) only, the chiral limit value of the $B^*B\pi$ coupling. 
At NLO and for local interpolating fields, two more LECs enter the results. One of them is known rather well and a second one entering the $B^*B\pi$ coupling is estimated by power counting for now. Thus ChPT makes concrete predictions for the $B\pi$-state contamination in the effective $B$-meson mass and decay constant and determines the order of magnitude for the summation estimate of the $B^*B\pi$ coupling. The extension to smeared 
interpolating fields requires the knowledge of the dependence of the LEC $\tilde \beta_1(r_\mathrm{sm})$ on the smearing radius $r_\mathrm{sm}$, which still needs to be computed from simulations.

A preliminary set of  results has already been given in \cite{Bar:2022jlw}.
%
\section{QCD with a static bottom quark}
%

\subsection{Action and symmetries}
In the following we consider QCD with a heavy $b$ quark. To LO in the heavy quark expansion (``static limit'') the Lagrangian in euclidean space time is given by \cite{Eichten:1989zv}
\bea\label{Lag_HQETStat}
{\cal L}&=&\sum_r \overline{q}_r (\gamma_{\mu} D_{\mu} + m_l) q_r+ \overline{Q} (D_4 + m_b) Q -\frac{1}{2g_0^2} \sum_{\mu\nu} \tr F_{\mu\nu}F_{\mu\nu} \,.
\eea
Here $q_r$ are the light up and down quark fields as well as possibly the strange and charm ones. The field $Q$ describes the heavy bottom quark taken to be at vanishing three-velocity.\footnote{We consider the theory in the $b$ quark's rest frame (RF) with $\vec{v}=0$, because this is all that is needed for our applications and also because the finite velocity theory is theoretically on less solid grounds \cite{Aglietti:1992in,Aglietti:1993hf}.} 
 The heavy field satisfies the constraint 
 \be 
 \label{e:constraint}
 Q=\gamma_4Q\,,
 \ee 
 thus it is a non-relativistic two-spinor. The contribution of the heavy anti-quark field is dropped since it
 is not needed in the following.

There is a rich  literature on heavy quark effective theory (HQET). For introductions to the subject we refer to \cite{Neubert:1993mb,Manohar:2000dt,Grozin:2004yc,Sommer:2010ic}  and the references therein. We here summarise a few properties that we will need later on. 

According to \eqref{Lag_HQETStat} the heavy quark propagates in time only. Its propagator is easily written down for an arbitrary gauge field and shows that the heavy quark mass $m_b$ only appears with an explicit factor $\exp(-m_b (x_4 -y_4)$). Therefore, $m_b$ shifts all energies of states with a single heavy quark. Keeping this in mind we may remove $m_b$ from the Lagrangian \pref{Lag_HQETStat} and add it to the energies later.

The static theory possesses various symmetries that are important for the construction of the corresponding chiral effective theory. 
First, the Lagrangian \pref{Lag_HQETStat} is invariant under the transformation
\be\label{TrafoQHQSS}
Q\,\rightarrow \,Q^{\prime} = S Q\,,\qquad \overline{Q}  \,\rightarrow \, \overline{Q}^{\prime} =\overline{Q} S^{\dagger}\,,
\ee
with $S$ being an SU(2) matrix satisfying $[\gamma_4,S]=0$. It corresponds to a rotation of the heavy quark spin (HQS) \cite{Isgur:1989vq,Georgi:1990um}.

Second, the static heavy quark Lagrangian is invariant under a local U(1) transformation that implies the conservation of so-called local flavor number (LFN) \cite{Sommer:2010ic},
\be\label{TrafoQLFN}
Q(x)\,\rightarrow \, Q^{\prime}(x) = e^{i \eta(\vec{x})} Q(x)\,,\qquad \overline{Q} \,\rightarrow \, \overline{Q}^{\prime}(x)=e^{-i \eta(\vec{x})}\overline{Q}(x) \,.
\ee
The invariance holds for arbitrary space (but not time) dependent functions $\eta(\vec{x})$. Both
\eqref{TrafoQHQSS} and \eqref{TrafoQLFN} hold in dimensional 
regularisation and in the lattice regularisations frequently used \cite{Kurth:2000ki}.

Third, we have the usual non-singlet chiral transformations, of the 
light quark fields, softly broken by the small masses of the light quarks. Separating the light quark fields in right-handed components, 
    $q_{R} =\frac{1+\gamma_5}{2} q$, and and left-handed ones, $q_{L} =\frac{1-\gamma_5}{2} q$, the symmetry transformations exlicitly read 
\begin{eqnarray}
&&q_{R} \, \rightarrow\, q_{R}^{\prime} = R q_{R}\,,\phantom{^\dagger} \qquad q_{L} \, \rightarrow\, q_{L}^{\prime} = L q_{L}\,, \\
&&\overline{q}_{R} \, \rightarrow \, \overline{q}_{R}^{\prime} = \overline{q}_{R}R^{\dagger} \,, \qquad \overline{q}_{L} \, \rightarrow \, \overline{q}_{L}^{\prime} = \overline{q}_{L}L^{\dagger}\,,
\end{eqnarray}
with $R,L$ being independent elements of the chiral symmetry groups SU$(N_f)_{R,L}$. The spontaneous breaking of this  symmetry is the basis of the chiral expansion \cite{Weinberg:1978kz,Gasser:1983yg,Gasser:1984gg}. While it might be  broken at the Lagrangian level in one form or another in lattice regularisations, it reappears in the continuum limit when the theory is properly renormalised~\cite{Bochicchio:1985xa}.

In the following we will be interested in correlation functions with interpolating fields for heavy $B$-mesons with a light anti-quark, i.e.\ either the $B^-$, $\overline{B}^0$ or the associated vector mesons. Furthermore, we need the flavor currents as we want to consider e.g. $b\to u$ transitions. We therefore first discuss static-light 
quark bilinears. We go a little into details, since we want to spell out the precise implications of the heavy quark symmetries listed above  including the 
effects of renormalisation and matching. The discussion is expected to hold beyond perturbation theory in the QCD coupling, see \cite{Sommer:2015hea} for some details.

\subsection{Static-light bilinears and their renormalization} \label{ssect:staticlightbilinears}

We discuss the full set of bilinears,
\be\label{bilin}
 J_r^\Gamma(x)= \overline{q}_r(x)\Gamma Q(x)\,,
\ee
with $\Gamma$ an arbitrary element of the Clifford algebra. In more common notation, we have in particular the vector and axial vector currents,
\bea
  V_{\mu,r} & = & J_r^{\gamma_\mu }, \qquad A_{\mu,r} \,=\, J_r^{\gamma_\mu \gamma_5},
\eea
and the scalar and pseudo scalar densities
 \bea
 S_{r} &=& J_r^{1}, \qquad P_{r} \,=\, J_r^{\gamma_5}.
\eea
The constraint \eqref{e:constraint} implies amongst others the  relations
\bea\label{VCPCPCAC}
 V_{4,r} &=&  S_r \,,\qquad
   A_{4,r} = - P_r \,.
   \eea
Note that space-time rotations are only a symmetry of the light 
sector of the theory.\footnote{Despite the fact that the 
$Q$-field Lagrangian breaks this symmetry, observables
of purely light quarks and gluons are symmetric under 
space-time rotations because the heavy quark determinant is trivial. Thus there is no influence of the heavy sector 
onto the light one. }
 Thus, $V_{k,r}$ is a vector and 
$A_{k,r}$ a pseudo-vector under 3-d rotations but the time-like fourth components are not related by 4-d rotations. Nevertheless,
we use the notation $A_k,A_4$ etc. 

\begin{figure}[t]
\begin{center}
\includegraphics[scale=0.55]{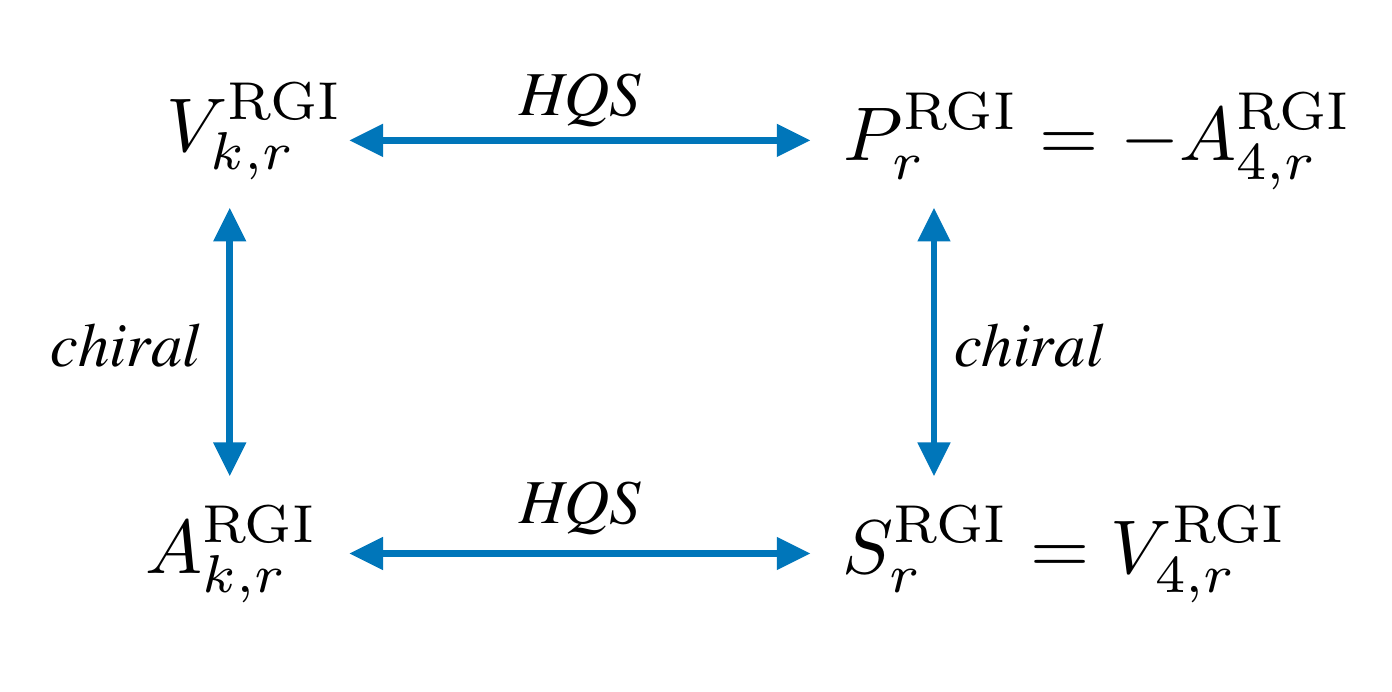}\\[0.4ex]
\caption{Symmetry relations of heavy-light bilinear renormalization group invariant currents.    
}
\label{fig:Illustration}
\end{center}
\end{figure}

Heavy quark symmetries and chiral symmetry of the light quarks introduce additional relations. In particular,
the special spin transformation with $S=R_k=\frac12 \epsilon_{ijk}\sigma_{ij}$ yields 
\be\label{SymTrafoHQS}
P_r \to {\cal R}_k(P_r) = V_{k,r}\,, \quad S_r \to {\cal R}_k (S_r)=A_{k,r} \,.
\ee
As an example this entails the equality (no summation over $k$)
\be
\langle P_r(x) P_r^\dagger(y) \rangle \,=\, \langle V_{k,r}(x) V_{k,r}^\dagger(y) \rangle\,.\label{SymPPVkVk}
\ee
Therefore, pseudo-scalar and vector heavy-light mesons 
are mass-degenerate in the static approximation, with a mass splitting that starts to appear at order $1/m_b$.
In addition, the special (finite) chiral rotation $\bar q \to  {\cal A}(\bar q)= -\bar q \gamma_5$
relates exactly the currents and the densities\footnote{This chiral SU(2) transformation is the combination of an axial transformation with $R_1=L_1^{\dagger}=i\sigma_3$ followed by a vector transformation with $R_2=L_2=-i\sigma_3$.},
\bea
V_{\mu,r} &\to& {\cal A}(V_{\mu,r}) = A_{\mu,r}\,, \quad A_{\mu,r} \to {\cal A}(A_{\mu,r}) = V_{\mu,r}\,,\label{SpecChirTrafoCurrents}\\
P_r &\to& {\cal A}(P_r) = - S_r\,, \qquad \,\,S_r \to {\cal A} (S_r)= - P_r\,.\label{SpecChirTrafoDensities}
\eea
Taking all symmetries together,  $A_{4}$ can be transformed by symmetry transformations into any $J_r^\Gamma$. Therefore, the renormalization,
\be
 J_{r,\mathrm{R}}^\Gamma(x;\mu) = Z_\mathrm{stat}(\bar g,\epsilon) \, J_{r}^\Gamma(x)\,,
\ee
here written in dimensional  regularisation with regularisation parameter $\epsilon$,
proceeds by a single common factor $Z_\mathrm{stat}$ for all bilinears.\footnote{We here have simplified a bit. Many regularisations including 
dimensional regularisation break chiral symmetry. 
It is understood that the symmetry is restored 
by imposing the chiral Ward identities. An exception 
is provided by lattice formulations with an exact 
lattice chiral symmetry~\cite{Luscher:1998pq}.} The
anomalous dimension $\gamma_\mathrm{stat}$ \cite{Voloshin:1986dir,Politzer:1988wp} in 
\be 
  \mu \frac{\rmd}{\rmd \mu} J_{r,\mathrm{R}}^\Gamma(x;\mu)=\gamma_\mathrm{stat}(\bar g)\, J_{r,\mathrm{R}}^\Gamma(x;\mu)\,, \quad \gamma_\mathrm{stat}(\bar g) =\gamma^\mathrm{stat}_0 \bar g^2 +\rmO(\bar g^4)\,,
  \quad \gamma^\mathrm{stat}_0=-\frac4{(4\pi)^2}\,,
\ee 
is then common to all $J_{r,\mathrm{R}}^\Gamma$  and renormalization group invariant currents are given by
\bea 
  J_{r,\mathrm{RGI}}^\Gamma(x) &=& \varphi_\mathrm{stat}(\bar g(\mu)) J_{r,\mathrm{R}}^\Gamma(x;\mu) \,,\label{RGICurrents}
  \\
  \varphi_\mathrm{stat}(g) &=& (2b_0 g^2)^{-\gamma^\mathrm{stat}_0/(2b_0)}\exp\{-\int_0^g \rmd x\, [\frac{\gamma_\mathrm{stat}(x)}{\beta(x)}- \frac{\gamma^\mathrm{stat}_0}{b_0 x}] \}\,,
\eea 
with one common $\varphi_\mathrm{stat}$. We use the conventions of \cite{Sommer:2010ic} for the renormalisation group functions. The scale-independent RGI bilinears are related by the symmetries discussed above, as illustrated in fig.\ \ref{fig:Illustration}. It is these bilinears that we will match to HMChPT in section \ref{sect:IntFieldsHMChPT}. 

\subsection{Bilinears matched to QCD}
Thus far we have assumed that the renormalisation conditions do not destroy 
the symmetries of the static theory. However, the fundamental theory, QCD, 
does not obey spin symmetry and we have to modify the
finite renormalisation of the effective theory such that it
reproduces QCD (up to the dropped $\rmO(1/m_b)$ corrections). 
We write the matched bilinears,
\be 
    J_{r,\mathrm{HQET}}^\Gamma(x) = C_\Gamma(M_b/\Lambda_{\msbar})\,J_{r,\mathrm{RGI}}^\Gamma(x)
\ee
in manifestly renormalisation group invariant terms. The scale- and scheme-invariant
RGI quark mass is defined as
\be
M_b=\lim_{\mu\to\infty} (2b_0 \bar g^2(\mu))^{-d_0/(2b_0)}\bar m_b(\mu)\,,
\ee
with $d_0=8/(4\pi)^2$. Chiral symmetry of massless QCD entails 
\be 
   C_\Gamma = C_{\gamma_5\Gamma}\,,
\ee
but the QCD interactions are not spin symmetric and therefore, e.g., $C_{\gamma_4\gamma_5} \ne  C_{\gamma_k}$. Following again the notation of \cite{Sommer:2010ic} we label the conversion functions by 
\be 
   C_\mathrm{PS} \, \equiv\, C_{\gamma_5\gamma_4}\,,\qquad 
   C_\mathrm{V} \,\equiv\, C_{\gamma_k}\,,
\ee
according to the particles, pseudo scalar (PS) and vector (V), which are
interpolated by the fields. The conversion 
function of the 
field $P_r$ is not independent. It is fixed by the 
partial conservation of the axial current,
\be 
\partial_\mu A_{\mu,r}^\mathrm{QCD} = (M_r + M_b) P_{r,\mathrm{RGI}}^\mathrm{QCD} \label{e:PCACQCD} \,.
\ee
Here the  current is assumed to be normalised canonically, 
such that the chiral Ward identities hold, and we have written 
the right hand side in terms of the RGI quantities, making
its scale invariance explicit. We now take the $B$-to-vacuum matrix element of the operator relation of the 
matched static fields, 
\be 
 C_\mathrm{PS}\,\langle 0 | \partial_4 A_{4,r}^\mathrm{RGI} | B(\vec p=0)\rangle = C_{{\gamma_5}} M_b \,\langle 0 |P_{r}^\mathrm{RGI}| B(\vec p=0)\rangle\,,
\ee
dropping $M_r$ which is suppressed by $\rmO(1/m_b)$.
Identifying $A_{4,r}^\mathrm{RGI} = -P_{r}^\mathrm{RGI}$ 
and noting $\langle 0 | \partial_4 P_{r}^\mathrm{RGI} | B(\vec p=0)\rangle= -M_B \langle 0 | P_{r}^\mathrm{RGI} | B(\vec p=0)\rangle$
with $M_B$ being the $B$-meson mass yields
\begin{equation}
C_{\gamma_5} = \frac{M_B}{M_b} C_\mathrm{PS} \,. 
\end{equation}
Note that taking the matrix element with a $B$-meson is a natural
choice, but not unique. Other choices would result in a change of 
the $M_B$ factor by a contribution suppressed by $1/m_b$. It is, however, not legitimate to set $M_B/M_b$ to one. That factor 
varies logarithmically with $m_b$.

Let us summarise the results of the preceding discussion 
which are relevant for the construction of HMChPT. The HQET bilinears $J_{r,\mathrm{HQET}}^\Gamma(x)$ are the ones which are matched to QCD. 
Their matrix elements yield the QCD matrix elements up to $\rmO(1/m_b)$. They obey  chiral symmetry and LFN conservation, but not heavy quark spin symmetry. The violation of this symmetry is encoded in
\be 
C_\mathrm{PS}/C_\mathrm{V} = 1 + [c_\mathrm{PS}-c_\mathrm{V}]  \alpha(m_b) + \rmO(\alpha(m_b)^2)\,.
\label{e:CPSV}
\ee
It is due to the spin-dependent interactions in QCD. A natural
choice for the heavy quark mass $m_b$ is $m_b^\star$ with $\overline m_{\msbar}(m^\star) = m^\star$. Independent of that choice,
\eqref{e:CPSV} has very large $\rmO(\alpha(m_b)^2)$ corrections \cite{Bekavac:2009zc} as illustrated e.g. in the appendix of \cite{Sommer:2010ic}. One thus has to use 
perturbative expansions of the matching functions with great care.

The relevant bilinears for the following are
\be 
   A_{4,r}^\mathrm{HQET} = C_\mathrm{PS} A_{4,r}^\mathrm{RGI}\,,
   \quad 
   V_{k,r}^\mathrm{HQET} = C_\mathrm{V} V_{k,r}^\mathrm{RGI}\,,
   \quad
   P_{r}^\mathrm{HQET} = -\frac{M_B}{M_b}\,C_\mathrm{PS}\, A_{4,r}^\mathrm{RGI}\,,  
\ee
with the RGI currents satisfying the relations
\bea 
V_{k,r}^\mathrm{RGI} &=& {\cal R}_k\left(A_{4,r}^\mathrm{RGI}\right)\,,\label{RGISymVkA4}\\
P_{r}^\mathrm{RGI} &=& -A_{4,r}^\mathrm{RGI}\,.\label{RGISymPA4}
\eea 
Furthermore, chiral symmetry allows to get the opposite parity
bilinears, in particular
\be 
V_{4,r}^\mathrm{RGI} = {\cal A}\left(A_{4,r}^\mathrm{RGI}\right)\,.\label{RGISymV4A4}
\ee 
Altogether, we need to construct only $A_{k,r}^\mathrm{RGI}$
in HMChPT with its associated low energy constants and all 
other bilinears are given by the above relations,
which are exact at leading order in $1/m_b$, which is the accuracy aimed for here. 
Below we will allow for the absence of chiral symmetry 
and also have $V_{k,r}^\mathrm{RGI}$ with independent 
low energy constants.
%
\subsection{Smeared interpolating fields}\label{ssect:smearedInterpolatingFieldsQCD}
%
Straightforward interpolating fields are the heavy-light pseudoscalar density, $P_r$ and the spatial components of the heavy-light vector current, $V_{k,r}$\,.
In practice, however,  so-called smeared interpolators are very often used to suppress excited-state contributions in correlation functions. These are formed as in \pref{bilin} but with the local anti-quark field $\overline{q}_r$ replaced with a smeared one $ \overline{q}^{\rm\, sm}_r$, which is generically of the form
\begin{equation}
\overline{q}^{\rm \,sm}_r (x) = \int {\rm d^4}y \, \overline{q}_r(y) K(y,x)\,,
\end{equation}
with some gauge covariant kernel $K(y,x)$ which decays rapidly for $|y-x|$ larger than some ``smearing radius'' $r_\mathrm{sm}$. The dependence of $K$ on the gauge field is suppressed. In the free theory it depends only on $x-y$ and not on $x,y$ separately. We denote the smeared currents, obtained by 
$\bar q_r \to \bar q_r^\mathrm{sm}$ in $J_r^\Gamma$ by $J_{r,\mathrm{sm}}^\Gamma$.

Different smearing procedures define different kernels. The gradient flow \cite{Luscher:2013cpa} is a four-dimensional, renormalisable smearing. 
The smearing of Ref.~\cite{Papinutto:2018ajw} is
local in time and the kernel is the propagator of a 
fermion on the time-slice. It is again renormalisable. The original
Gaussian and exponential (scalar) smearing \cite{Gusken:1989ad,Gusken:1989qx,Alexandrou:1990dq} are local in time. Following the discussion of \cite{Papinutto:2018ajw} we conclude that the scalar smearing is {\em not} renormalisable while the relation of the Gaussian smearing with a 3-d fermion gradient flow  \cite{Luscher:2013cpa} restricted to a time slice suggests that it might be renormalisable. 

In the following the transformation properties of the smeared interpolating fields will be relevant. 
The horizontal symmetries in Fig.~\ref{fig:Illustration}, are properties of
the static field $Q$. They are obviously satisfied 
irrespective of the smearing type.\footnote{The chiral symmetry relations can be enforced for the 
{\em renormalisable} smearings mentioned before, since they 
are based on a local formulation (with the help of an extra flow-time dimension~\cite{Luscher:2013cpa} or 
an extra fermion field~\cite{Papinutto:2018ajw}).  Then the smeared anti-quark fields transform just as the local ones under all the symmetries sketched in Fig.~\ref{fig:Illustration}.} Consequently, also the fields $J_{r,\mathrm{sm}}^\Gamma$ transform under HQS just as their local counterparts. This  property essentially determines their expressions in ChPT,  see section \ref{sect:IntFieldsHMChPT} for the explicit construction through NLO. 
%
\section{$B\pi$ excited-state contamination in correlation functions}\label{sect:BpiInCorrFunctions}
%
\subsection{2-point function and the $B$-meson mass and decay constant}\label{ssect:2ptfunction}

Throughout this article we consider QCD/HQET in a finite spatial box.  $L$ denotes the box length in  each direction and boundary conditions are assumed such that meson fields are periodic in space. The euclidean time extent, however, is taken infinite. This choice implies a simple exponential decay of $n$-point functions. This setup slightly simplifies our calculations and the infinite volume limit can be taken in the end, if desired. 
Another simplification concerns the masses of the up and down type quarks which we assume to be equal. 

We are interested in the two-point (2-pt) correlation function of a $B$-meson interpolating field ($t>0$),
\begin{eqnarray}\label{DefC2pt}
C_{2,rr'}(t)& =& \int_{L^3} {\rm d}^3{{x}}\, \langle  B_r(\vec{x},t) B_{r'}^{\dagger}(0,0)\rangle \equiv \delta_{rr'}C_{2}(t).
\end{eqnarray}
The field $B_r$ stands for the renormalized RG invariant pseudoscalar density $ J_{r,\mathrm{RGI}}^{\gamma_5}$ in \pref{RGICurrents}, or for the smeared density involving a spatially smeared light anti-quark. For simplicity only we switch to the more intuitive notation $B_r$. Because of the symmetry \pref{SymPPVkVk}  we do not need to consider the vector current 2-pt function.  

The integration over the spatial volume in Eq.~\pref{DefC2pt} projects on states with zero total momentum. 
Performing the standard spectral decomposition in \pref{DefC2pt} the 2-pt function is found to be a sum of various contributions,
\begin{equation}\label{C2specdecompGen}
C_{2}(t) = C^B_{2}(t) + C^{B\pi}_{2}(t) + \ldots \,.
\end{equation}
The first term on the r.h.s. stems from the single-heavy-meson (SHM) state with the $B$-meson being at rest,\footnote{If the smearing procedure involves a smearing in time the simple exponential decay is modified at distances of the order of the smearing radius.} 
\begin{equation}
\label{spcontr}
C_{2}^{B}(t)=
\frac{1}{2}\;|\langle 0|B(0)|B(\vec{p}=0)\rangle|^2 e^{-M_{B} t } \,.
\end{equation}
Here 
$|B(\vec{p}=0)\rangle$ denotes the $B$-meson state with the meson at rest and we assume this state to be normalized according to the non-relativistic condition in a finite volume,
\begin{equation}
\langle B(\vec{p})|B(\vec{p}^{\,\prime})\rangle \,=\, 2L^3 \delta_{\vec{p},\vec{p}^{\,\prime}}\,.
\end{equation}

The interpolating field excites other states with the same quantum numbers as the $B$-meson. For sufficiently small pion masses and large volumes the dominant multi-hadron states are those containing additional pions. For the two-particle $B\pi$ contribution\footnote{HQSS relates $B$ and a $B^{\ast}$ exactly. Therefore, we generically speak of $B\pi$ states even if the two-particle excited state involves a $B^{\ast}$.}  in the spectral decomposition we find
\begin{eqnarray}\label{tpcontr}
C_{2}^{B\pi}(t)&=&\frac{1}{L^3}\;\sum_{\vec{p}}\frac{1}{4 \Ep}\,
|\langle 0|B(0)|B(-\vec{p}) \pi(\vec{p} )\rangle|^2 e^{-E_{B\pi}t}\,.
\end{eqnarray}
Here $E_{B\pi}$ is the total energy of the two-particle state. Its normalisation is taken over from the limiting case $|B(-\vec{p}) \pi(\vec{p} )\rangle \to  |B(-\vec{p})\rangle |\pi(\vec{p} )\rangle$ where the two particles do not interact. We use the standard relativistic normalisation for the pion states, $\langle \pi(\vec{p})|\pi(\vec{p}^{\,\prime})\rangle \,=\, 2E_{\pi,\vec{p}}L^3 \delta_{\vec{p},\vec{p}^{\,\prime}}$. Indeed, for small $p=|\vec p|$ the pions interact weakly with the $B$-mesons and  $E_{B\pi}$ equals approximately the sum $E_{B,\vec{p}}+\Ep$, with 
\begin{equation}
\Ep=\sqrt{p^2+M_{\pi}^2}\,,\quad E_{B,\vec{p}}=M_B+\rmO(1/M_B)\,.
\end{equation}
In \pref{tpcontr} the sum over  momenta runs over all momenta allowed by the boundary conditions imposed for the finite spatial volume, i.e.\ $ \vec{p}=2\pi\vec{n}/L$ with $\vec{n}$ having integer-valued components in case of periodic boundary conditions.

The two-particle $B\pi$ contribution is smaller than the SHM contribution. On one hand there is the exponential suppression with the pion energy and the time separation $t$. At least in principle it can be made arbitrarily small by choosing $t$ sufficiently large. In practice, however, the signal-to-noise problem \cite{Parisi:1983ae,Lepage:1989hd} limits the accessible time separations to $t\lesssim 2$ fm.  In addition to the exponential suppression we expect the prefactor involving the inverse spatial volume and the ratio of the two matrix elements to be small too. The latter can be estimated from the following argument. The matrix elements on the rhs of eqs.\ \pref{spcontr} and \pref{tpcontr} have mass dimension 3/2 and 1/2, respectively. For fixed $m_\pi$, we can therefore write their ratio as
\begin{equation}\label{Ratio}
  \frac{\langle 0|B(0)|B(-\vec{p}) \pi(\vec{p})\rangle}{ \langle 0|B(0)|B(0)\rangle} = \frac1{f_\pi}F(p/M_\pi) = \kappa \frac{p}{f_\pi M_\pi} + 
  \frac1{f_\pi}\rmO(p^2/M_\pi^2),
\end{equation}
with a dimensionless function $F$ and a dimensionless constant $\kappa$. 
A $\vec{p}$-independent term is forbidden in the expansion of $F$ by parity conservation.
Therefore, we may conclude that
\begin{equation}\label{NaiveEst}
\frac{C^{B\pi}_2(t)}{C^{B}_2(t)}\approx  \sum_{\vec{p}}\,\frac{\kappa^2}{2(f_{\pi}L)^2\Ep L}\,\frac{{p}^{2}}{M_\pi^2}\,\big(1+\rmO(p^2/M_{\pi}^2)\big)\,e^{-\Ep\, t}\,,
\end{equation}
and our explicit HMChPT computation in section \ref{sect:BpiInHMChPT} shows that $\kappa$
is smaller than one.
If we assume the values $M_{\pi} = 180$ MeV and $L = 5$ fm we roughly find $[2(f_{\pi} L)^2 E_{\pi}L]^{-1}\approx 1/80$ for the smallest non-vanishing pion momentum $p=2\pi/L$. 
However, the smallness of this estimate must not be mistaken with the size of the entire $B\pi$ contribution, since many $B\pi$ states contribute to the sum. In particular, the number of states relevant in the sum increases rapidly with the spatial volume, leaving a non-vanishing contribution in the infinite volume limit.

Multi-particle-states with more than one pion contribute to
\pref{C2specdecompGen} analogously to \pref{tpcontr}, but each additional pion contributes an additional factor $[2(f_{\pi} L)^2\Ep L]^{-1}e^{- \Ep t}$, i.e.\ the more pions in the state the larger the suppression.

The $B\pi$ contribution in the 2-pt function enters directly the effective $B$-meson mass commonly employed in numerical lattice simulations to measure the $B$-meson mass. Defined as $M_B^{\rm eff}(t) = -\partial_t \ln C_2(t)$ we approximately find
\begin{equation}\label{EstEffBmesonMass}
M_B^{\rm eff}(t) \approx M_B + \kappa^2\sum_{\vec{p}}\frac{\Ep}{2(f_{\pi}L)^2\Ep L}\frac{p^2}{M_\pi^2}e^{-\Ep\, t}\,.
\end{equation}

Our discussion so far holds for rather arbitrary interpolating fields, and in particular for the local heavy-light axial vector current. For this special case the matrix element in \pref{spcontr} is essentially the $B$-meson decay constant,
\begin{eqnarray}\label{DefMEA4}
\langle 0|A_4^\mathrm{RGI}(0)|B(\vec{p}=0)\rangle  &\equiv& \fhat = [C_\mathrm{PS}(M_b/\Lambda_{\overline{\rm MS}})]^{-1}f_B \sqrt{M_B}  \,.
\end{eqnarray}
A simple estimator for this observable is given by \cite{Sommer:2010ic}
\begin{eqnarray}\label{DefFhatestimator}
\fhatsqreff(t) &=& 2 C_2(t) e^{M_B^{\rm eff}(t)  \,t},
\end{eqnarray}
which tends to \pref{DefMEA4} for $t\rightarrow \infty$. For finite $t$ this estimator too inherits the excited-state contamination in $C_2(t)$ and $M_B^{\rm eff}(t)$. 

While the estimator in \pref{DefFhatestimator} works in principle it is hardly ever used in practice since the excited state contamination in it is typically very large. An alternative estimator for $\hat{f}$ (not the square) is given by
\begin{eqnarray}\label{DefFhatestimatorLS}
\fhateff(t) &=& \sqrt{2} \frac{C^{LS}_2(t)}{\sqrt{C^{SS}_2(t)}} e^{\frac{1}{2}M_B^{\rm eff}(t)  \,t}\,.
\end{eqnarray}
Here $C^{LS}_2(t)$ refers to the 2-pt function with the local $A_4$ at the sink and a smeared interpolating field $B^{\dagger}$ at the source. Analogously, $C^{SS}_2(t)$ denotes the 2-pt function with smeared interpolators at both source and sink. Employing at least one smeared interpolating field in the 2-pt function leads to a significantly smaller excited-state contamination in the estimator for the $B$-meson decay constant~\cite{Alexandrou:1990dq}.

The estimate for the $B\pi$ contribution in \pref{NaiveEst} stems from a simple dimensional analysis and Taylor expansion in $|\vec{p}|$ for the matrix elements of the $B$-meson interpolation field. It can be put on more solid grounds by employing ChPT. In section \ref{sect:BpiInHMChPT} we compute the ratio in \pref{NaiveEst} to NLO in the chiral expansion.

\subsection{3-point function and the $B^*B\pi$-coupling}

Excited $B\pi$ states enter observables other than the effective mass as well. A prominent example is the $B^*B\pi$ coupling that is commonly extracted from a suitably defined 3-pt function.
This coupling is defined as the matrix element of the light-light axial vector current between a pseudoscalar and a vector $B$-meson state,\footnote{Note that the definition involves the matrix element in Minkowski space, as indicated by the upper index $j$ on the lhs. Note also that the charged axial vector components $A^{\pm}=A^1 \pm i A^2$ lead to a matrix element that is twice as large compared to the neutral component with $a=3$. }
\begin{equation}\label{DefgPi}
\langle \bar{B}^0(0)|A^{j,-}(0)|B_{k}^{*,-}(0)\rangle = 2 g_{\pi} \delta_{jk}\,.
\end{equation}
To obtain this matrix element we consider the euclidean 3-pt function (no sum over $k$)
\begin{eqnarray}\label{DefC3}
C_3(t,t') &=& \int_{L^3} d^3x \int_{L^3} d^3y \,\langle \bar{B}^0(t,\vec{x}) A_k^{-}(t',\vec{y}) {B_k^{*,-}}^\dagger(0,\vec{0})\rangle\,
\end{eqnarray}
and the ratio with the 2-pt function,\footnote{The matrix element in euclidean space-time involves an additional factor $i$ \cite{Best:1997qp}, which is taken into account in the definition of the ratio. 
} 
\begin{equation}\label{DefRatio}
R(t,t') \equiv - i \frac{C_3(t,t')}{C_2(t)}\,.
\end{equation}
Performing the spectral decomposition of the two correlation functions and taking all times $t, t'$ and $t-t'$ to go to infinity, it is straightforward to show that the ratio goes to a constant. The prefactor $-i$ is chosen such that this constant is the desired coupling,
\begin{equation}
R(t,t')\, \longrightarrow\, g_{\pi}\,.
\end{equation}

For finite times $t,t'$ the ratio involves contributions due to excited states.  Generically, it can be written as
\begin{equation}
R(t,t') = g_{\pi}\Big(1+ \Delta R^{B\pi}(t,t')+\ldots \Big)\,
\end{equation}
where $ \Delta R^{B\pi}(t,t')$ denotes the excited state contribution due to $B\pi$ states, and the ellipses stand for all other excited state contributions. The $B\pi$ contribution is of the form (we always assume $t > t' >0$ unless explicitly specified otherwise)
\begin{equation}\label{DefDeltaBpi}
\Delta R^{\!B\pi}(t,t') = \sum_{\vec{p}}\left( b(\vec{p})\,[ e^{-\Ep (t-t')} +  e^{-\Ep\, t'}] + c(\vec{p}) e^{-\Ep\, t}\right)\,.
\end{equation}
We can distinguish two types of $B\pi$ contributions: (i) the ground-state-to-excited-state contribution, captured by the coefficient $b\propto \langle B^*\pi|A|B^*\rangle = \langle B|A|B \pi \rangle$ (the two matrix elements are related by heavy quark spin symmetry together with time reversal symmetry), and (ii) the excited-state-to-excited-state contribution given by the coefficient $c$. The latter involves the matrix element $\langle B^*\pi|A|B\pi\rangle$ stemming from the 3-pt function, but also contains the $B\pi$ contribution in the 2-pt function.  
In section \ref{sect:BpiInHMChPT} we compute these coefficients through NLO in ChPT.
 
Based on the ratio $R(t,t')$ there are two popular estimators for $\gpi$. The first one is the {\em mid-point estimate}, defined by
 \begin{equation}\label{Defmidpointestimator}
\gmid = R(t,t/2)\,.
\end{equation}
It is motivated by the observation that the excited-state contribution, for a given source-sink separation $t$, is minimal if the operator is placed in the middle between source and sink. 
The second one is the {\em summation estimate} \cite{Maiani:1987by,Capitani:2012gj}, given by 
\begin{equation}\label{Defsummestimate}
\gsum = \frac{d}{dt}\int_0^t dt'\, R(t,t')\,.
\end{equation}
Both estimators inherit the $B\pi$ state contribution from the ratio $R$, and are parameterised by the same coefficients $b,c$. 
Explicitly, we find for the two estimators the expressions
\begin{eqnarray}
\label{e:gmid}
\gmid &=& \gpi\left( 1+ \sum_{\vec{p}} \, \Big[ 2b(\vec{p}) e^{-\Ep \,t/2 } + c(\vec{p}) e^{-\Ep\, t } \Big]\right)\,,\\
\label{e:gsum}
\gsum & = & \gpi\left( 1+ \sum_{\vec{p}} \, \Big[ 2b(\vec{p}) + c(\vec{p}) (1 - \Ep\, t)  \Big] e^{-\Ep\, t } \right)\,,\label{gPisumest}
\end{eqnarray}
in terms of the coefficients introduced before. Once we have the ChPT results for these coefficients at hand we obtain estimates for the impact of $B\pi$ excited states on the estimators for the $B^*B\pi$-coupling.
%
\section{Heavy meson chiral perturbation theory}\label{sect:HMChPT}
%
\subsection{Preliminaries}
The correlation functions introduced in the last section can be computed in ChPT. Provided the euclidean time separations are sufficiently large the correlators are dominated by pion physics and ChPT is expected to provide reliable results for them. 

ChPT, the low-energy effective theory of QCD, is based on the spontaneous breaking of chiral symmetry, leading to light pseudo Nambu-Goldstone bosons, the pions. Their interaction with themselves and with other particles are heavily constrained by chiral symmetry and its breaking by the quark masses. Coupling the pions to heavy quark mesons, so-called Heavy-Meson (HM) ChPT has been introduced a long time ago \cite{Wise:1992hn,Burdman:1992gh,Yan:1992gz}. For reviews we refer the reader to Refs.\ \cite{Wise:1993wa,Grinstein:1994au,Casalbuoni:1996pg}, for example.

In the following we collect some basic definitions and results that are necessary in later sections. The main motivation is to settle our notation since we work, following Ref.\ \cite{Bernardoni:2009sx}, 
in euclidean space time. In addition, we work in the RF of the $B$-meson and display our definitions and results for this particular choice only.

\subsection{Particle content}

The doublets of pseudoscalar and vector $B$-mesons are $({B}^-, \bar{B}^0)$ and $({B}^{*-}, \bar{B}^{*0})$, respectively. In the effective theory they are described by complex pseudoscalar and vector fields. To keep the notation simple we denote the pseudoscalar doublet by $P$ and the vector mesons by $P_k,\, k=1,2,3$, suppressing  the light flavor index $r=1,2$ and the asterisk for the vector mesons. Both $P, P_{k}$ are complex valued fields, and in the following we use the asterisk $*$ to denote complex conjugation, and $\dagger$ for the matrix or vector adjoint. 

To manifestly preserve the heavy quark spin symmetry it is common and convenient to combine the fields $P,P_{k}$ and their complex conjugates in  multiplets $H,\bar{H}$ defined by \cite{Wise:1992hn}
\bea
H & = & P_+\left( i P_{k} \gamma_{k} + i P \gamma^5\right)\,,\label{DefHMHfield}\\
\bar{H} & = & \gamma_4 H^{\dagger} \gamma_4 \, = \,  \left(i P^{\dagger}_{k} \gamma_{k} +i P^{\dagger} \gamma_5\right)P_+\,,\label{DefHMHbarfield}
\eea
with the projector $P_+= (1+\gamma_4)/2$.\footnote{Our definition for $H$ differs from the one in Ref.\ \cite{Bernardoni:2009sx} by an overall sign, but this 
 sign is irrelevant in the Lagrangian since it contains both fields $H$ and $\bar{H}$, see below.}
 The original fields $P,P_{k}$ are recovered from $H$ by a "projection" according to
\bea\label{projH}
 \langle H\gamma_5\rangle & =& 2 i P\,, \qquad  \langle H\gamma_{k}\rangle = 2i P_{k}\,,
\eea
where $\langle\cdots\rangle \equiv {\rm tr\,}[\cdots]$ denotes the trace over the $\gamma$-matrix indices. 
Note that the field $H$ satisfies 
\begin{equation}\label{PropHg4}
H \, = \, \gamma_4 H\,  =\,  - H \gamma_4\,,
\end{equation}
a property that will be useful later on. 

The field $H$ represents the heavy $B$-meson with valence quark content $Q\overline{q}$. Therefore, $H$  inherits its transformation properties under HQS and LFN symmetries from the heavy quark field $Q$, see \pref{TrafoQHQSS} and \pref{TrafoQLFN}. Explicitly we require
\bea
H & \longrightarrow & S H \,,\qquad \,\, {H} \, \longrightarrow\, e^{i\eta(\vec{x})}{H}\,,\label{HQSandLFNtrafoH}\\
\bar{H} & \longrightarrow  &  \bar{H}S^{\dagger} \,,\qquad \bar{H} \, \longrightarrow \, \bar{H}e^{-i\eta(\vec{x})}\,.
\eea
In ChPT the Goldstone boson fields are typically collected in the matrix valued field
\be\label{DefU}
U = \exp\left({\frac{2 i}{f}\pi}\right) \,,\qquad \pi = \pi^aT^a\,,
\ee
with real valued pseudoscalar fields $\pi^a, a=1,2,3$ and the SU(2) group generators $T^a=\sigma^a/2$. 
When coupled to heavy particles the square root of $U$ is the natural field,
\be
u = \sqrt{U} = \exp\left({\frac{ i}{f}\pi}\right)\,.
\ee
Under chiral transformations $R,L$  this field and its complex conjugate transform as
\bea
u & \longrightarrow & Lu h^{\dagger} = h u R^{\dagger}\,,\label{ChiralSymu}\\
u^{\dagger} & \longrightarrow & Ru^{\dagger} h^{\dagger} = h u^{\dagger}  L^{\dagger}\,,\label{ChiralSymudag}
\eea
where $h$ is the usual compensator field depending on $R,L$ and the pion fields. $u,u^{\dagger}$ are used to form the combinations 
\bea
u_{\mu} & = &\frac{i}{2} \{u^{\dagger} (\partial_{\mu} - i l_{\mu}) u - u (\partial_{\mu} - i r_{\mu}) u^{\dagger}\}\,,\label{buildblocks1a}\\
\chi_{\pm} & =& u^{\dagger} \chi u^{\dagger} \pm u\chi^{\dagger} u\,.\label{buildblocks1b}
\eea
Here $r_{\mu}=v_{\mu} + a_{\mu}$, $l_{\mu}=v_{\mu} - a_{\mu}$ and $\chi = 2B (s+i p) $ contain the external source fields for the light quark vector and axial vector currents as well as the scalar and pseudoscalar density. The former ones transform as gauge fields under local chiral transformations \cite{Gasser:1983yg,Gasser:1984gg}. Under chiral transformations  the combinations in \pref{buildblocks1a}, \pref{buildblocks1b} transform the same way,
\bea
u_{\mu} &\longrightarrow &h u_{\mu} h^{\dagger}\,,\label{Def:umu}\\
\chi_{\pm}& \longrightarrow &h \chi_{\pm} h^{\dagger}\,.\label{Def:chi}
\eea
The compensator field $h$ also appears in the transformation laws for the heavy meson fields,
\bea
P& \longrightarrow & P h^{\dagger}\,,
\qquad
P_{k} \, \longrightarrow \, P_{k}  h^{\dagger}\,,
\label{ChiraltrafoPmu}
\eea
therefore the multiplets $H,\bar{H}$ transform analogously,
\bea\label{ChiraltrafoH}\
H & \longrightarrow H h^{\dagger} \,,\qquad \bar{H} \longrightarrow h \bar{H}\,.
\eea
The invariance under local chiral transformations implies that derivatives acting on the heavy meson fields must be covariant derivatives that preserve the transformation of the fields. It is given by $D_{\mu} = \partial_{\mu} -i \Gamma_{\mu}$  with the connection
\bea
\Gamma_{\mu} = \frac{i}{2}\{u^{\dagger} (\partial_{\mu} - i l_{\mu}) u + u (\partial_{\mu} - i r_{\mu}) u^{\dagger}\}\,.\label{Def:Gammamu}
\eea
The covariant derivative acts on the heavy meson fields $P, P^{\dagger}$ as
\bea
D_{\mu} P = \partial_{\mu} P +i P \Gamma_{\mu}\,,\qquad D_{\mu} P^{\dagger} = \partial_{\mu} P^{\dagger} -i  \Gamma_{\mu} P^{\dagger}\,,
\eea
and the same way on the vector field. By construction, the covariant derivatives of the fields transform as in \pref{ChiraltrafoH}.
\subsection{Chiral Lagrangian}\label{ssect:chiralLag}

\begin{figure}[t]
\begin{center}
\includegraphics[scale=0.16]{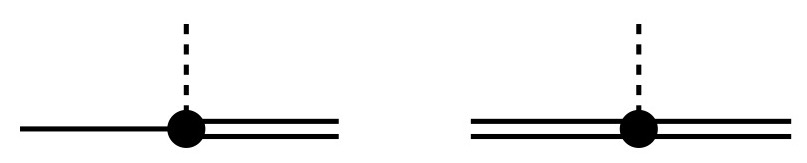}\\[0.4ex]
\caption{
Two interaction vertices in ${\cal L}^{(1)}_{\rm HM}$, coupling a pion with a heavy pseudoscalar and vector meson (left) and with two heavy vector mesons. Both vertices are proportional to the coupling $g$.  
}
\label{fig:int_vertices}
\end{center}
\end{figure}
The chiral effective Lagrangian relevant for the application in this paper is the sum of three terms, 
\begin{eqnarray}
{\cal L}_{\rm eff} & = & {\cal L}^{(1)}_{\rm HM}+  {\cal L}^{(2)}_{\rm HM} + {\cal L}^{(2)}_{\pi}\,.
\end{eqnarray}
Here ${\cal L}^{(2)}_{\pi}$ denotes the standard mesonic Lagrangian for the Goldstone bosons \cite{Gasser:1983yg,Gasser:1984gg},
\bea\label{L2PureMesonic}
{\cal L}^{(2)}_{\pi} & =& {f^2} \langle u_{\mu} u_{\mu} \rangle - \frac{f^2}{2} \langle \chi_+ \rangle\,.
\eea
We write it  using the combination $u_{\mu}$ defined in \pref{buildblocks1a}, but it is easily checked that it coincides with the standard form expressed in terms of $U=u^2$. $f$ denotes the pion decay constant in the chiral limit, and our notation corresponds to $f\approx 93$ MeV.

The LO heavy meson Lagrangian consists of two terms only,
\bea\label{LOHMLag}
{\cal L}^{(1)}_{\rm HM} & = & -  \langle D_{4} H  \bar{H}\rangle - i g  \langle H \gamma_{5} \gamma_{\mu} u_{\mu} \bar{H} \rangle\,.
\eea 
Recall that $H$ is a row vector as far as flavor symmetry is concerned. Thus $\bar{H}$ is a column vector and the sum over the flavor indices is kept implicit in \pref{LOHMLag}. 
The two interaction terms in \pref{LOHMLag} contain one  derivative (of the pion field), as indicated by the superscript on the left hand side in eq.\ \pref{LOHMLag}. 

The equations of motion (EOM) for the fields $H$ and $\bar{H}$ can be derived as usual and we find
\begin{eqnarray}
i D_{4} H & = & g H \gamma_{5} \gamma_{\nu} u_{\nu}\,.\label{EOMH}
\end{eqnarray}

\begin{table}[bt]
\begin{center}
\begin{tabular}{c|l}
\hline
\hline
$\,\,k\,\,$ & $\,\,\,O^{(2)}_k$\\
\hline
1 & $\,\,\,\langle H u_{\mu} u_{\mu} \bar{H}\rangle$ \\
2 & $\,\,\, \langle H u_{4} u_{4}  \bar{H}\rangle $\\
3 & $\, i \langle H u_{\mu} u_{\nu} \sigma_{\mu\nu} \bar{H}\rangle $\\
4 & $\,\,\,\langle H f_{+,\mu\nu} \sigma_{\mu\nu} \bar{H}\rangle $\\
5 & $\,\,\,\langle H \langle f_{+,\mu\nu}\rangle  \sigma_{\mu\nu} \bar{H}\rangle \,\,$\\
6 & $\,\,\,\langle H \chi_+ \bar{H}\rangle $\\
7 & $\,\,\,\langle H \langle \chi_+\rangle  \bar{H}\rangle $\\
\hline\hline
\end{tabular}
\caption{\label{table:NLOtermsLag} The terms of chiral dimension 2 for SU(2) HMChPT in the static limit as listed in \cite{Jiang:2019hgs}. }
\end{center}
\end{table}%

The Lagrangian in \pref{LOHMLag} contains an infinite number of interaction terms with an arbitrary number of pion fields. Relevant in the following are  the ones with only one pion field that are contained in the term proportional to the coupling $g$. 
They are obtained by replacing $u_{\mu}$with $- \partial_{\mu}\pi /f$ in \pref{LOHMLag}. The two vertices we obtain are displayed in figure \ref{fig:int_vertices}. Their explicit expressions for the Feynman rules are given in appendix \ref{app:Feynman_rules}, together with the propagators stemming from the first term in \pref{LOHMLag}.

The NLO Lagrangian ${\cal L}^{(2)}_{\rm HM}$ involves seven terms in case of SU(2) HMChPT in the static limit \cite{Jiang:2019hgs},
\begin{eqnarray}\label{NLOHMLag}
{\cal L}^{(2)}_{\rm HM} & = & \sum_{k=1}^7 d_k  O^{(2)}_k\,,
\end{eqnarray}
with the individual $O^{(2)}_k$ listed in table \ref{table:NLOtermsLag}. In case of SU(3) the total number of terms increases to eight. 
$f_{+,\mu\nu}$ refers to a field combination involving the field strength tensor build from the external source fields for the light vector and axial vector currents, see Ref.\ \cite{Jiang:2019hgs} and  appendix  \ref{app:Feynman_rules}. It vanishes if the source fields are set to zero, thus the terms with $k=4,5$ do not contribute any interaction vertices.
In fact, none of the terms in table \ref{table:NLOtermsLag} spawns an interaction vertex relevant for our application. 
It is straightforward to convince oneself that the interaction terms in all $O^{(2)}_k$ involve an even number of pion fields, therefore, through NLO they do not contribute to the two-particle $B\pi$ contribution in the correlation functions we are interested in.

However, ${\cal L}^{(2)}_{\rm HM}$ does contribute at NLO to the light axial vector current that we need for the 3-pt function defined in \pref{DefC3}. Recall that the light axial vector current can be obtained by a derivative of the chiral action with respect to the source field $a_{\mu}(x) = a^c_{\mu}(x) T^c$,
\bea\label{Def:lightAmu}
A^{c}_{\mu}(x) & = &\left. \frac{\delta {S}_{\rm eff}[a_{\mu}]}{\delta a_{\mu}^c(x)}\right|_{a_{\mu}=0}\,.
\eea
Here  LO and NLO terms stem from ${\cal L}^{(1)}_{\rm HM}$ and ${\cal L}^{(2)}_{\rm HM}$, respectively, and both provide a non-vanishing contribution. An additional contribution originates in the mesonic part ${\cal L}^{(2)}_{\pi}$.
The explicit expressions are collected in appendix \ref{app:Feynman_rules}. Figure \ref{fig:axial_vect_vertices} shows the resulting vertices for the Feynman diagrams. The LO vertex is proportional to the coupling $g$. At NLO there are terms with LECs $d_k$ stemming from \pref{NLOHMLag}.  
More details are given in appendix \ref{app:Feynman_rules}. 

\begin{figure}[bt]
\begin{center}
\includegraphics[scale=0.4]{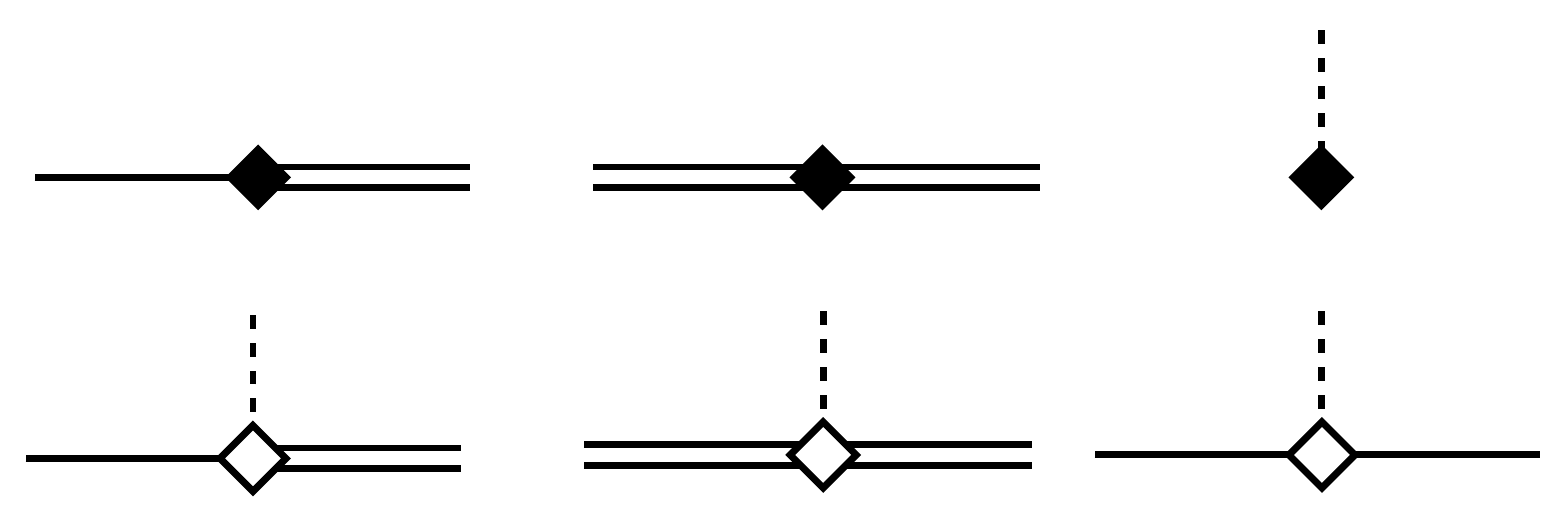}\\[0.4ex]
\caption{
The vertices for the axial vector current, represented by the diamond. The full symbol in the top row stand for the vertices stemming from  ${\cal L}^{(1)}_{\rm HM}$ and ${\cal L}^{(2)}_{\pi}$. The open symbol in the bottom row represent the NLO vertices stemming from  ${\cal L}^{(2)}_{\rm HM}$.   
}
\label{fig:axial_vect_vertices}
\end{center}
\end{figure}

%
\section{Interpolating fields for the heavy $B$ mesons in HMChPT}\label{sect:IntFieldsHMChPT}
%
\subsection{Preliminaries}

In order to compute the correlation functions introduced in section \ref{sect:BpiInCorrFunctions} within HMChPT we still need the HMChPT expressions for the interpolating fields for the $B$-mesons. To our knowledge these expressions are not given in the literature, thus we provide the construction through NLO in this section.

The construction is standard and is based on the general principles used to formulate ChPT: We write down the most general expressions that transform the same way under the various symmetry transformations as the expressions in the underlying theory, here the static limit of QCD. The  principle to order the terms is, as usual,  the increasing chiral dimension of the individual terms, i.e.\ the number of derivatives 
or the number of light quark masses. Each independent term in this expansion is associated with an a priori unknown LEC. 

It will be useful and illustrative to be rather general in the following. We will construct the HMChPT expressions of the full set of currents and densities introduced in section \ref{ssect:staticlightbilinears}. As discussed, the RGI bilinears exhibit a high degree of symmetry relating the various currents and densities, e.g.\ \pref{RGISymVkA4} -- \pref{RGISymV4A4}. This high degree of symmetry manifests itself in a small number of independent LECs in the chiral effective theory.\footnote{ Note that the HMChPT expressions for the HQET bilinears are obtained by a  multiplication with factors $C_{\rm V}$, $C_{\rm PS}$ and $M_B/M_b$.}
In a second step we consider the bilinears with a smeared light quark field (smeared bilinears for short). 

\subsection{Constructing the currents and densities in HMChPT}\label{ssect:constructingCurrents}

We start with the construction of the HMChPT expressions for the local heavy-light vector and axial vector currents.
We first construct individual vector and axial vector components $O_{V}$ and $O_{A}$ that transform appropriately under chiral transformations and parity.  To automatically guarantee the proper transformation behaviour under the LFN symmetry, cf.\ \pref{HQSandLFNtrafoH}, we consider expressions with one field $H$ only. The actual spatial components of the vector and axial vector currents 
are obtained with the usual projection by taking the trace over the $\gamma$-matrix indices,
\bea\label{projectionOmu}
V_{k} = \langle O_{V}\gamma_{k} \rangle\,,\qquad A_{k} = \langle O_{A} \gamma_{k} \rangle\,,
\eea
for $k=1,2,3$.
In ChPT one expands in powers of pion momenta and quark masses. Following the notation in Ref.\ \cite{Wein:2011ix} we write
\bea 
O_X = \sum_{n=0}^{\infty} O_X^{(n)} \,,\quad X\,=\,V,A\,.
\eea
The superscript $n$ denotes the chiral dimension of $O_X^{(n)}$, and every $O_X^{(n)}$ itself is a finite sum of various terms, 
\begin{eqnarray}\label{defORLn}
O_{X}^{(n)} &=& \sum_{j=1}^{N_n}  \alpha^{(n)}_{j;X} O_{j;X}^{(n)} \,,\quad X\,=\,V,A\,.
\end{eqnarray}
Here $N_n$ denotes the number of independent terms for every $n$. The $\alpha^{(n)}_{j;X}$ are the LECs associated with the individual terms.
Note that the LECs for general vector and axial vector currents are not related until we impose chiral symmetry, see below.

HQS symmetry relates the scalar and pseudo scalar densities to the spatial components of the currents. In HMChPT this is achieved 
if the $O_{V},O_{A}$ are used in the projection to get the expressions for the pseudo scalar and scalar densities,
\bea\label{projectionO5}
P& =&  \langle O_{V}\gamma_{5} \rangle\,,\qquad S\, =\, -  \langle O_{A}\gamma_{5} \rangle\,.
\eea
Finally, the time-like components of the currents are defined such that 
the relations \pref{VCPCPCAC} are satisfied, i.e.\ we simply set
\begin{eqnarray}\label{DefFourCompCurr}
A_{4} & = &- P \,,\qquad V_{4} = S\,
\end{eqnarray}
with the expressions in \pref{projectionO5}. 

With \pref{projectionO5} and \pref{DefFourCompCurr} all we need to construct are the chiral expressions \pref{defORLn} for the spatial components of the currents.
Table \ref{table1} lists all terms $O_{j;V}^{(n)},O_{j;A}^{(n)}$ of chiral dimension $n=0$ (LO) and 1 (NLO), and two examples for terms of chiral dimension 2.\footnote{We omit terms that vanish identically when the projection \pref{projectionOmu} is performed.} 
For the construction note that $u,u^{\dagger}$ are the only independent building blocks that transform with a $R^{\dagger},L^{\dagger}$ on the right hand side, cf.\ \pref{ChiralSymu} and \pref{ChiralSymudag}. Consequently,  the terms for the vector and axial vector parts must end with $(u\pm u^{\dagger})$ on the right hand side. $u,u^{\dagger}$ can be combined with $u_{\mu}$ or $\chi_{\pm}$, but this increases the chiral dimension. Lorentz indices must be  properly contracted among two $u_{\mu}$ or with an open index provided by a Clifford algebra element. 

\begin{table}[t]
\begin{center}
\begin{tabular}{|c|c|c|c|}
\hline\hline
$\,\,n\,\,$ & $\,\,j\,\,$ &  $O^{(n)}_{j,V}$& $O^{(n)}_{j,A}$  \rowheight \\
\hline
0 & 1 &   $H(u+ u^{\dagger})$& $H(u- u^{\dagger})$ \\
\hline
1 & 1 & {\rule[-0mm]{0mm}{4.5mm}} $\,\,H \gamma_{k} \gamma_{5} u_{k}(u+u^{\dagger})\,\,$ & $\,\,H \gamma_{k} \gamma_{5} u_{k}(u-u^{\dagger})\,\,$ \\[0.4ex]
1 & 2 &  $H u_{4}(u-u^{\dagger})$ & $H  u_{4}(u+u^{\dagger})$ \\[0.4ex]
1 & 3 &  $i D_{4} H (u + u^{\dagger})$ & $i D_{4} H (u - u^{\dagger})$ \\[0.8ex]
\hline
2& 1&   $H \chi_+ (u +u^{\dagger})$&$H \chi_+ (u-u^{\dagger})$ {\rule[-0mm]{0mm}{4.5mm}} \\[0.2ex] 
2& 2&   $H u_k u_k (u +u^{\dagger})$&$H u_k u_k (u-u^{\dagger})$ {\rule[-0mm]{0mm}{4.5mm}} \\[0.2ex] 
$\vdots$ & $\vdots$ & $\vdots$ & $\vdots$ \\[0.8ex]
\hline
\hline
\end{tabular}
\end{center}
\caption{\label{table1} List of $O^{(n)}_{j,X}$, $X=V_k,A_k, V_4, A_4$,  for the lowest $n$.}
\end{table}

For $n=0$ and $1$ table \ref{table1} lists two and six terms, respectively. Note that we have two independent combinations, $u_k \gamma_{k}$ and $u_4\gamma_4$ since Lorentz symmetry is broken. In the second combination we can drop $\gamma_4$ since the heavy meson field satisfies eq.\ \pref{PropHg4}. For the same reason there are no $n=1$ terms involving $\sigma_{k 4}$. The LFN symmetry forbids terms with a derivative acting on the heavy meson field $H$, except for  $i D_{4}$.

By construction, the currents and densities automatically satisfy the HQS symmetry. Chiral symmetry, on the other hand, is not satisfied unless the LECs satisfy some constraints.
The special finite chiral rotation leading to \pref{SpecChirTrafoCurrents}, \pref{SpecChirTrafoDensities} is given in the chiral effective theory by $u\rightarrow h u$, $u^{\dagger} \rightarrow - h u^{\dagger}$. Hence, it transforms the entries of the two columns in table \ref{table1} into each other, i.e.\ $O^{(n)}_{j,V}\leftrightarrow O^{(n)}_{j,A}$.  
Consequently, the symmetry relation \pref{SpecChirTrafoCurrents} is satisfied in the chiral effective theory if 
\begin{equation}
\alpha^{(n)}_{j;V} = \alpha^{(n)}_{j;A}\,,  
\end{equation}
thus we can drop the label $X$ and simply write  $\alpha^{(n)}_{j} $.

Beyond this factor 1/2, the number of independent LECs can be further reduced by invoking the EOM. With eq.\ \pref{EOMH} we see that the $n=1$ terms with $j=3$ and $j=1$ are not independent.  We choose to drop the terms with $j=3$ in the following.

To summarize, imposing all symmetry constraints and with the use of the EOM we end up with one unknown LEC at LO and two unknown LECs at NLO for the entire set of currents and densities.   

As explicit results we give the expressions for the time-component of the axial current $A_4 = -P$ and the spatial components of the vector current $V_k$, valid to first order in the chiral expansion,
\begin{eqnarray}
A_4 &  =& -\frac{\alpha}{2} P u_+ + i\frac{\beta_1\alpha}{2} P_k u_k u_+ - i\frac{\beta_2\alpha}{2} P u_4 u_- \,,\label{A4Expl}\\
V_k & =& \frac{\alpha}{2} P_k u_+ + i \frac{\beta_1\alpha}{2} (Pu_k - \epsilon_{klm} P_l u_m) u_+ + i \frac{\beta_2\alpha}{2} P_k u_4 u_- . \label{VkExpl}
\end{eqnarray}
To simplify the notation we have introduced the short hand notation
\begin{eqnarray}\label{RedefLECs}
\alpha \equiv 4i\alpha^{(0)}_1\,\qquad \beta_1\alpha \,\equiv\, 4 \alpha^{(1)}_{1} \,,\qquad \beta_2\alpha \,\equiv\, -4\alpha^{(1)}_2\,.
\end{eqnarray}

\subsection{Smeared interpolating fields in HMChPT}\label{ssect:smearedInterpolatingFields}

So far we have considered the pointlike currents and densities and matched them to HMChPT. 
In practice smeared interpolating fields, $J_{\mathrm{sm}}^{\gamma_5} = P^\mathrm{sm}$ 
and $J_{\mathrm{sm}}^{\gamma_k} = V_{k}^\mathrm{sm}$ are used 
in order to suppress excited state contaminations. 
We now map them to effective fields in HMChPT and use $\tilde{\phantom{\alpha}}$ to refer to the ChPT expressions for smeared quark bilinears, e.g.\    
$\tilde{O}_{X}^{(n)}$ in the expansion \pref{defORLn} or 
$\tilde{P}$ and $\tilde{V}_{k}$ for the smeared pseudoscalar and vector meson interpolating fields.

As discussed above, $P^\mathrm{sm}$ and $V_k^\mathrm{sm}$ are related exactly by HQS symmetry at any fixed lattice spacing and for the same smearing kernel. 
This symmetry is preserved as long as we use the same $\tilde{O}_{V}$ in the projection on the smeared pseudoscalar and vector meson interpolating fields,
$\tilde{P} = \langle \tilde{O}_{V}\gamma_{5}\rangle$ and $\tilde{V}_{k} = \langle \tilde{O}_{V}\gamma_{k} \rangle$. Since we only need $\tilde{O}_{V}$, the same LECs, entering the expansion 
\begin{eqnarray}\label{defOtildeV}
\tilde O_{V}^{(n)} &=& \sum_{j=1}^{N_n}  \tilde{\alpha}^{(n)}_{j, V} {O}_{j;V}^{(n)}\,,
\end{eqnarray}
appear in both the HMChPT expressions for the smeared pseudoscalar and vector meson interpolating fields.
Therefore, the expansion \pref{defOtildeV} involves one LEC at LO, and two at NLO.
Note that we do not (and do not need to) impose the chiral symmetries for this conclusion. This is beneficial since the chiral symmetries  are compromised in one way or the other in most lattice formulations. 

The explicit results for the smeared interpolators $\tilde{V}_k, \tilde{A}_4 = - \tilde{P}$ are as in \pref{A4Expl} and \pref{VkExpl} once the LECs are properly replaced, ${\alpha}^{(n)}_{j, V}\rightarrow \tilde{\alpha}^{(n)}_{j, V}$. In analogy to \pref{RedefLECs} we introduce the corresponding short hand notation
\begin{eqnarray}\label{RedefTLECs}
\tilde{\alpha} \equiv 4i\tilde{\alpha}^{(0)}_1\,\qquad \tilde{\beta}_1\tilde{\alpha} \,\equiv\, 4 \tilde{\alpha}^{(1)}_{1} \,,\qquad \tilde{\beta}_2\tilde{\alpha} \,\equiv\, -4\tilde{\alpha}^{(1)}_2\,.
\end{eqnarray}

In general, the LECs $\tilde{\alpha}_{j,V}^{(n)}$ depend on 
the smearing radius $r_\mathrm{sm}$ and all the parameters of the lattice discretisation including the lattice spacing $a$, or equivalently the bare coupling $g_0$. For renormalisable smearings (see above), they are of a more restricted form $\tilde \alpha^{(n)}_{k}
= \eta(g_0) \bar \alpha^{(n)}_k(r_\mathrm{sm})$. This will be relevant later, since a common factor such as $\eta(g_0)$ drops out in the physical matrix elements. 
 The expressions for excited state contaminations will depend only on $r_\mathrm{sm}$ (discretisation errors are always neglected).

In mapping smeared interpolating fields to ChPT an additional condition has to be satisfied. Smeared interpolators with some ``size'' are mapped onto pointlike fields in the chiral effective theory. For this to be a good approximation the smearing radius needs to be small compared to the Compton wave length of the pion $r_{\rm sm} \ll M_{\pi}^{-1}$. For physical pion masses this is about 1.4 fm, and for smearing radii of a few tenths of a fermi the bound is reasonably well satisfied. 

\begin{figure}[t]
\begin{center}
\includegraphics[scale=0.16]{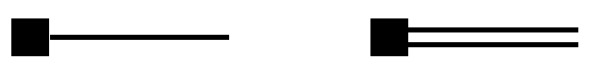}\\[0.4ex]
\includegraphics[scale=0.16]{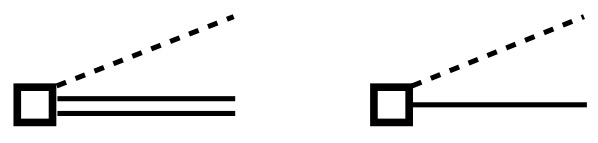}\\[0.4ex]
\caption{
Interaction vertices for the smeared interpolating fields, on the left for the pseudo scalar, on the right for the vector $B$ meson. The LO vertices (top row) involve the LO LEC $\tilde{\alpha} \equiv 4i \talo{1}$, the NLO ones (bottom row) come with the LEC $\tilde{\beta}$, a combination of the NLO LECs $\tilde{\alpha}^{(1)}_j$. The Feynman rules for these vertices are collected in appendix \ref{app:Feynman_rules}. 
}
\label{fig:IntField_vertices}
\end{center}
\end{figure}

\subsection{Comment: The left-handed heavy-light current}

A by-product of our previous discussion is the chiral expression for the local left-handed current, which is obtained with the difference $O_L^{(n)} \,=\, O_V^{(n)} -O_A^{(n)}$. Explicitly we obtain the LO expressions
\begin{eqnarray}
L_{k}^{(0)} &=& 2i\alpha_1^{(0)}  \langle H  \gamma_{k} \rangle u^{\dagger}\,,\quad L_{4}^{(0)} \,=\, 2i\alpha_1^{(0)}  \langle H  \gamma_{5} \rangle u^{\dagger}\,.
\end{eqnarray}
This can be conveniently written as 
\begin{eqnarray}\label{LOLmu}
L_{\mu}^{(0)} &=& -\frac{i\alpha}{2} \langle H  \gamma_{\mu} (1-\gamma_5) \rangle u^{\dagger}\,,
\end{eqnarray}
where we used  $\langle H \gamma_k\gamma_5\rangle = \langle H \gamma_4 \rangle = 0$, and spatial and temporal components are related due to HQS, not by Lorentz symmetry. This suggestive form was first given by Wise in \cite{Wise:1992hn,Wise:1993wa} and it is commonly used in the literature, e.g.\ in Refs.\ \cite{Booth:1994hx,Sharpe:1995qp,Becirevic:2002sc,Aubin:2005aq,Bijnens:2010ws}. 

With the $n=1$ terms in table \ref{table1} we analogously obtain the NLO result
\begin{eqnarray}
L_{\mu}^{(1)} &=& \phantom{+}2 \anlo{1}  \langle H \slashed{u}  \gamma_{\mu} (1-\gamma_5) \rangle u^{\dagger}\nn\\
&&- 2 \anlo{2}  \langle H \gamma_{\mu} (1-\gamma_5) \rangle  u_{4} u^{\dagger}\nn\\
&& + 2 \anlo{3}  \langle iD_{4} H \gamma_{\mu} (1-\gamma_5) \rangle u^{\dagger}\label{Lmudim1}
\end{eqnarray}
in case no use is made of the EOM. To our knowledge the complete current at NLO, \pref{Lmudim1}, has not been constructed systematically in the literature.  However, Ref.\ \cite{Aubin:2005aq} quotes two terms as representative examples. It happens that the third term in the complete result \pref{Lmudim1} can be dropped by invoking the EOM and the terms given as examples in Ref.\ \cite{Aubin:2005aq} are in fact complete on-shell.

For the interpretation of the LO LEC $\alpha$ we compute the matrix element on the lhs in \pref{DefMEA4} to LO in ChPT.  Using $A_{4}=- P$  we find
\be\label{Identifikalpha}
\alpha \,=\,  \hat{f}\,,
\ee
i.e.\ the LEC $\alpha$ is the $B$-meson decay constant in the combined static and chiral limit. 

The decay constant has been calculated to one-loop in Refs.\ \cite{Sharpe:1995qp,Becirevic:2002sc,Aubin:2005aq}. For this calculation the LO current  \pref{LOLmu} is needed, together with parts of $L_{\mu}^{(2)} $ that provide the counter terms necessary to renormalise the result. The NLO current \pref{Lmudim1} is not needed for the result in \cite{Sharpe:1995qp,Becirevic:2002sc}. It contributes only at higher order in the chiral expansion. 

%
\section{$B\pi$ excited-state contamination in HMChPT}\label{sect:BpiInHMChPT}
%
\subsection{Preliminaries}
The 2-pt and 3-pt functions in eqs.\ \pref{DefC2pt}, \pref{DefC3} can be computed order by order in the chiral expansion.  Provided the time scales $t$ and $t'$ are large the correlation functions are dominated by pion physics and expected to be well-captured by HMChPT. 

Given the explicit expressions in the last section the perturbative expansion of the correlation functions is standard and straightforward. 
We find it convenient to work in a finite spatial volume with periodic boundary conditions, reflecting the setup commonly encountered in Lattice QCD. 
The explicit expressions for the propagators and vertices are summarized in appendix \ref{app:Feynman_rules}. 
The infinite volume limit of the results is easily obtained from the finite volume results. For simplicity we assume an infinite time extent leading to a simple exponential decay of the correlation functions. 

\subsection{2-point function and the $B$-meson mass}\label{ssect:Results2ptfunction}

\begin{figure}[t]
\begin{center}
\includegraphics[scale=0.35]{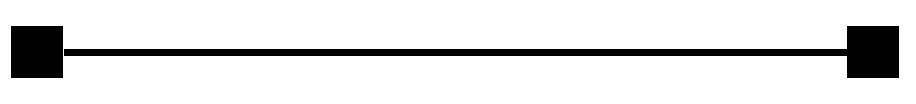}\\[0.4ex]
\caption{
LO Feynman diagram for the SHM contribution to the 2-pt function. The solid line represents the pseudo scalar $B$-meson propagator. The filled squares denote the LO interpolating fields at $t$ and $0$. 
}
\label{fig:diag_C2_SHM}
\end{center}
\end{figure}

Figure \ref{fig:diag_C2_SHM} shows the Feynman diagram that provides the leading SHM contribution to the 2-pt function, and the result reads (we always assume $t>0$)
\begin{eqnarray}\label{resC2BSHM}
C^B_2(t) = \frac{\tilde\alpha_{\rm sk} \tilde{\alpha}_{\rm src}}{2} e^{-M_B \,t}\,,
\end{eqnarray}
where $\tilde{\alpha}_{\rm sk}$ and $\tilde{\alpha}_{\rm src}$ are the LO LECs of the interpolating $B$-fields at sink and source.

\begin{figure}[t]
\begin{center}
\includegraphics[scale=0.35]{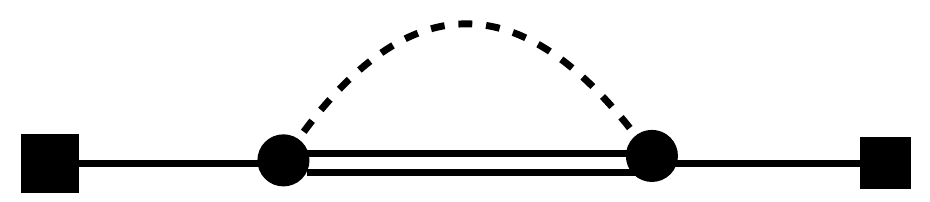}\hspace{1cm} \includegraphics[scale=0.35]{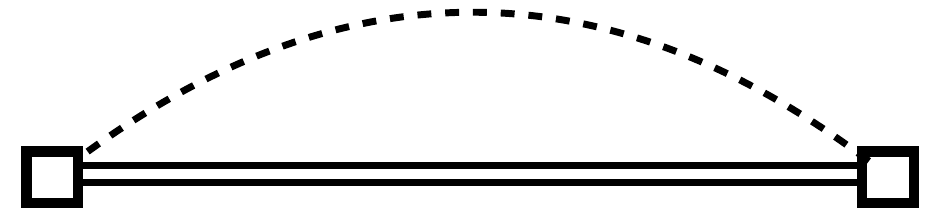}\\[0.4ex]
a)\hspace{4cm} b)\\[2ex]
\includegraphics[scale=0.35]{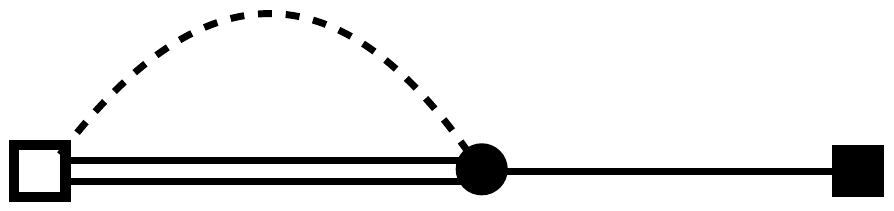}\hspace{1cm} \includegraphics[scale=0.35]{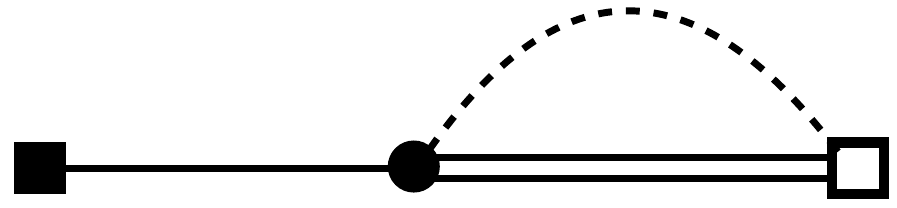}\\[0.4ex]
c)\hspace{4cm} d)\\[2ex]
\caption{
Feynman diagrams contributing to the $B\pi$ contribution through NLO. The dashed and the solid double line represent the pion and vector $B$-meson propagators, respectively, and the open squares denote the interpolating field at NLO. The LO interaction vertex is given by the filled circle and implies an integration over space-time. 
}
\label{fig:diags_C2_Bpi}
\end{center}
\end{figure}

The diagrams depicted in figure \ref{fig:diags_C2_Bpi} contribute to the $B\pi$ contribution $C^{B\pi}_2$. To quote the results it is useful to introduce some notation. We rewrite eq.\ \pref{C2specdecompGen} according to 
\begin{equation}\label{DefDeltaC2}
C_{2}(t) = C^B_{2}(t)\left(1 + \Delta C_2^{B\pi}(t) \right) \,,
\end{equation}
with $ \Delta C_2^{B\pi}(t) =  C_2^{B\pi}(t)/C_2^B(t)$. This ``deviation'' from the SHM contribution is of the general form
\begin{equation}\label{DefDeltaC2b}
 \Delta C_2^{B\pi}(t) = \sum_{\vec{p}} c_{\rm 2pt}(\vec{p}) \, e^{-\Ep\,t}\,,
\end{equation}
and, as discussed in section \ref{ssect:2ptfunction}, the generic structure of the coefficient $c_{\rm 2pt}$  is
\begin{equation}\label{DefC2ptCoeff}
c_{\rm 2pt}(\vec{p}) = U(\vec{p})\,C_{\rm 2pt}(\vec{p})
\,,\; U(\vec{p})=
\frac{3}{8 (fL)^2 \Ep L } \frac{p^2}{\Ep^2} \, \,.
\end{equation}
The universal function $U$ carries the entire volume dependence.
In contrast to eq.\ \pref{NaiveEst} we have included a factor $3/4$ here. This factor is the Casimir operator of the SU(2) flavor group, and it naturally appears in the computation. The non-trivial result of our ChPT calculation is the ``reduced'' coefficient $C_{\rm 2pt}(\vec{p})$, a dimensionless function of the pion momentum $\vec{p}/\Lambda_\chi$. 

The explicit calculation of the diagrams in fig.\ \ref{fig:diags_C2_Bpi} results in
\begin{equation}\label{resC2pt}
C_{\rm 2pt}(\vec{p}) = \left(g + \tilde{\beta}_{1,{\rm sk}}\Ep \right)\left(g + \tilde{\beta}_{1,{\rm src}} \Ep\right)\,.
\end{equation}
where $\tilde{\beta}_{1,{\rm sk}}$ and $\tilde{\beta}_{1,{\rm src}}$ are the low energy constants of the interpolating $B$-fields at sink and source.
Note that this is the ChPT result through NLO. The LO contribution is given by the term proportional to $g^2$, while the NLO correction is given by the term proportional to the LEC $\tba$. 

The LECs associated with the interpolating fields appear in the combination $\tba = -i \tilde{\alpha}_1^{(1)}/\tilde{\alpha}_1^{(0)}$. In case of pointlike interpolating fields we obtain the same result \pref{resC2pt} but with $\tba$ replaced by $\ntba=-i {\alpha}_1^{(1)}/{\alpha}_1^{(0)}$, the LEC combination associated with the pointlike interpolators. 
This appearance of these ratios will always be the case for the quantities studied in this paper, and it is the reason for having defined the NLO LECs as multiples of the LO LEC in \pref{RedefLECs} and \pref{RedefTLECs}.

The LEC $\ntba$ has mass dimension $-1$. The standard
assumption is that such an LEC is roughly of a magnitude
given by the inverse of the chiral symmetry breaking scale $\Lambda_{\chi}=4\pi f \approx 1{\rm GeV}$, and in this case we recover the standard chiral expansion in powers of $\Ep/\Lambda_{\chi}$ with dimensionless coefficients which are expected to be of O(1). 
This argument is slightly more complicated for smeared interpolating fields, since a second scale is present, the inverse smearing radius $r_{\rm sm}$. In this case we expect a chiral expansion with two expansion parameters.
If $r_{\rm sm}$ is substantially larger than $\Lambda_{\chi}^{-1} \sim 0.2$ fm the additional expansion parameter $\Ep\, r_{\rm sm}$ is larger than the original one  $\Ep/\Lambda_{\chi}$, potentially leading to a poorer chiral expansion. 
 
Note that the contributions of the four diagrams in fig.\ \ref{fig:diags_C2_Bpi} combine to \pref{resC2pt}. With the same smearing at source and sink, it is manifest that $C_{\rm 2pt}$ is a positive number in agreement with the general positivity of coefficients in the spectral decomposition \pref{C2specdecompGen}. This is the reason why we prefer to keep the $\Ep^2$ contribution in \pref{resC2pt}, even though formally it is higher than NLO.

The results \pref{DefDeltaC2b} to \pref{resC2pt} were derived for a finite spatial volume $L^3$. 
Taking the infinite volume limit using $\Delta p\,=\, {2\pi}/{L}$ and $\sum_p (\Delta p)^3 \,\rightarrow \int d^3p$\,,
we obtain the expression
\begin{equation}\label{Eq:DeltaC2_Inf_Vol}
\Delta C_2^{B\pi}(t) \,\rightarrow\, \frac{3}{16\pi^2f^2} \int_{M_{\pi}}^\infty d E\, \,\frac{p(E)^3}{E^2}   \left( g +\tba E \right)^2 e^{-E\, t}\,,
\end{equation}
with $p(E)=\sqrt{E^2-M_{\pi}^2}$ being the spatial pion momentum for a given energy $E=\Ep$. Carrying out the energy-integral leads
to Bessel- and Struvefunctions as detailed in App.~\ref{app:AnalyticIVresults}.

In eq.\ \pref{EstEffBmesonMass} we introduced the effective $B$-meson mass and derived the general form for the $B\pi$ contribution in it. Its deviation from the true mass is
\begin{eqnarray}\label{Res_DMeff}
\Delta M_B^{B\pi}(t) \equiv M_B^{\rm eff}(t) - M_B = M_{\pi} \sum_{\vec{p}} c_{\rm 2pt}(\vec{p}) \frac{\Ep}{M_{\pi}}\, e^{-\Ep\,t}\,
\end{eqnarray}
in terms of the coefficients introduced in \pref{DefC2ptCoeff}. The deviation $\Delta M_B^{B\pi}$ is dimensionful and we have chosen to express it as the pion mass times a dimensionless number given by the sum in \pref{Res_DMeff}. 
Here too the infinite volume limit can be easily taken, which leads to essentially the same expression as on the right hand side of \pref{Eq:DeltaC2_Inf_Vol} but with an additional factor $E/M_{\pi}$ in the integrand.

Finally, the $B\pi$ contamination in the effective decay constant \pref{DefFhatestimatorLS} stems from the ones in the 2-pt functions $C_2^{LS}, C_2^{SS}$ and the effective mass. Defining the deviation $\Dfhateff$ by
\begin{eqnarray}
\fhateff(t) &=& \fhat \left(1 + \Dfhateff(t)\right)\,,
\end{eqnarray}
we obtain 
\begin{eqnarray}\label{Res_Deltafhat}
\Dfhateff(t) &=& \Delta C_2^{LS,B\pi}(t) -\frac{1}{2} \left(\Delta C_2^{SS,B\pi}(t) - t\,\Delta M_B^{B\pi}(t)\right) \,.
\end{eqnarray}
Note that this result depends on two NLO LECs, $\tba$ and $\ntba$, the latter LEC stemming from the pointlike axial vector current $A_4$. 

\subsection{3-point function and the $B^*B\pi$-coupling}

\begin{figure}[t]
\begin{center}
\includegraphics[scale=0.35]{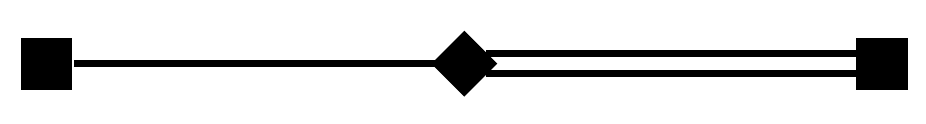}\\[0.4ex]
\caption{
LO Feynman diagram for the SHM contribution to the 3-pt function. The diamond represents the LO axial vector current at operator insertion time $t'$, the other elements are as in fig.\ \ref{fig:diags_C2_Bpi}. 
}
\label{fig:diag_C3_SHM}
\end{center}
\end{figure}

The computation of the $B\pi$ contribution in the 3-pt function \pref{DefC3} and the ratio $R$ in \pref{DefRatio} is analogous to the one in the 2-pt function.

The Feynman diagram in figure \ref{fig:diag_C3_SHM} yields the leading SHM contribution to the 3-pt function,
\begin{equation}
C^B_3(t,t') = i g \frac{\tilde\alpha_{\rm sk} \tilde{\alpha}_{\rm src}}{2}\,.
\end{equation}
Together with the SHM result \pref{resC2BSHM} for the 2-pt function we obtain the LO result $R(t,t') = g$. Consequently, for both the mid-point and summation estimator we obtain the LO results $\gmid=\gsum=g$.

\begin{figure}[tp]
\begin{center}
\includegraphics[scale=0.35]{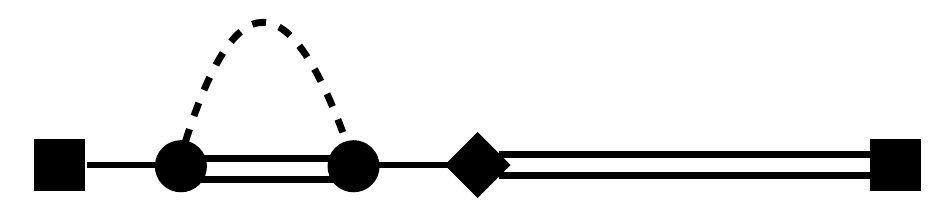}\hspace{0.5cm}\includegraphics[scale=0.35]{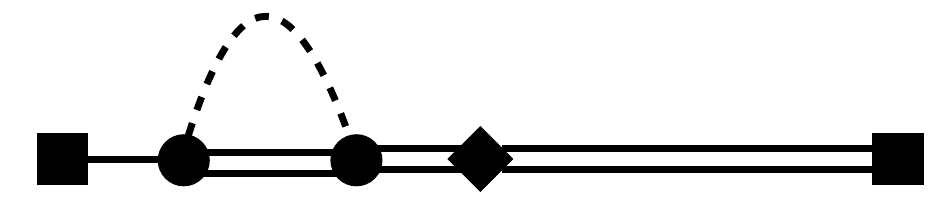}\hspace{0.5cm}\includegraphics[scale=0.35]{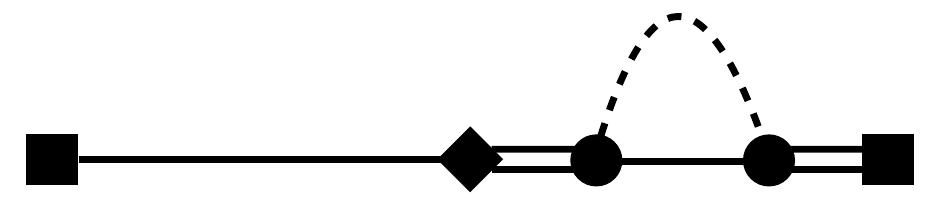}\hspace{0.5cm}\includegraphics[scale=0.35]{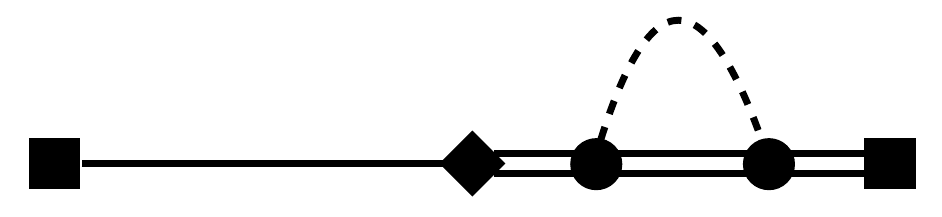}\\[0.4ex]
a)\hspace{3.3cm} b) \hspace{3.3cm} c)  \hspace{3.3cm} d)\\[2ex]
\includegraphics[scale=0.35]{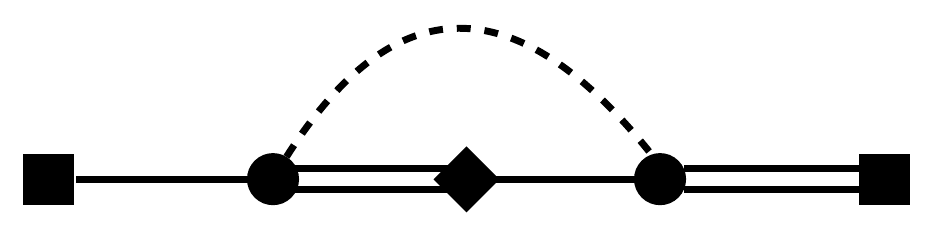}\hspace{0.5cm}\includegraphics[scale=0.35]{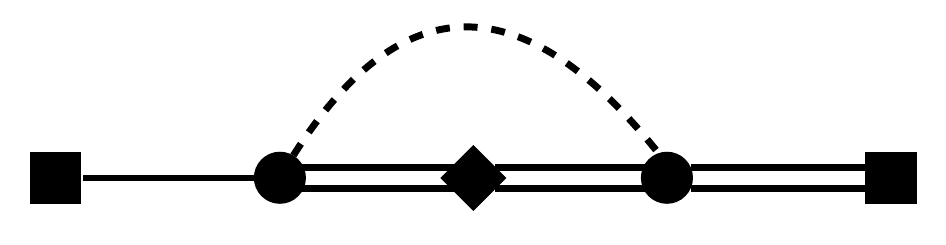}\hspace{0.5cm}\includegraphics[scale=0.35]{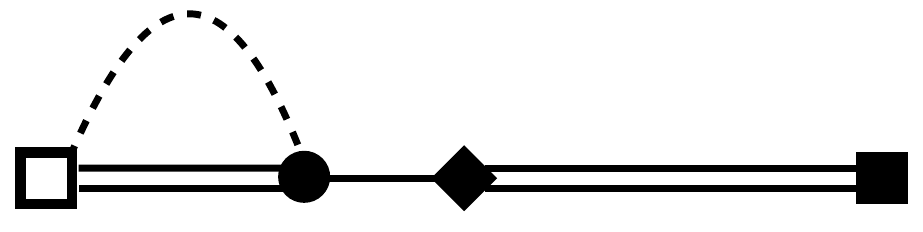}\hspace{0.5cm}\includegraphics[scale=0.35]{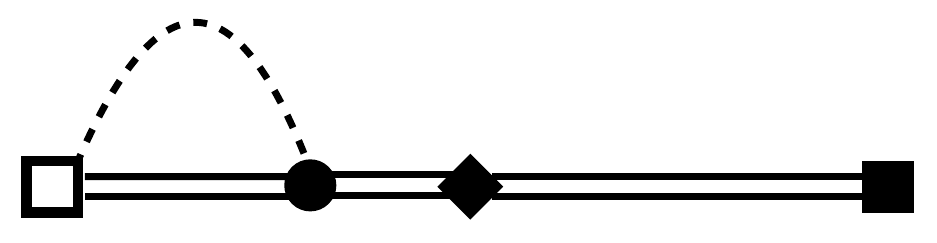}\\[0.4ex]
d)\hspace{3.3cm} e) \hspace{3.3cm} f)  \hspace{3.3cm} g)\\[2ex]
\includegraphics[scale=0.35]{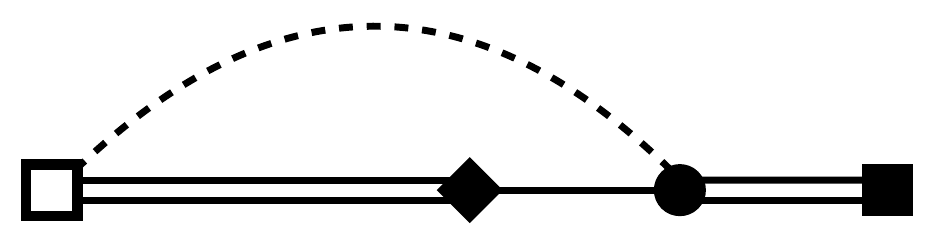}\hspace{0.5cm}\includegraphics[scale=0.35]{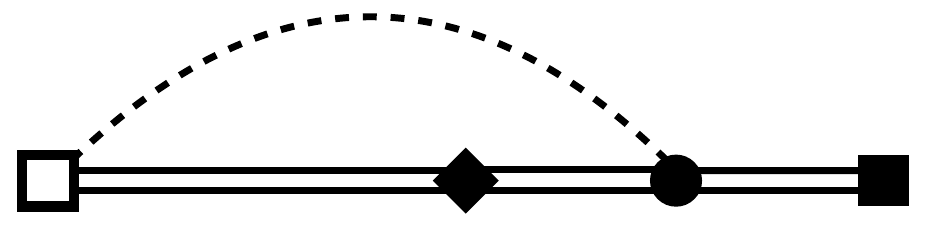}\hspace{0.5cm}\includegraphics[scale=0.35]{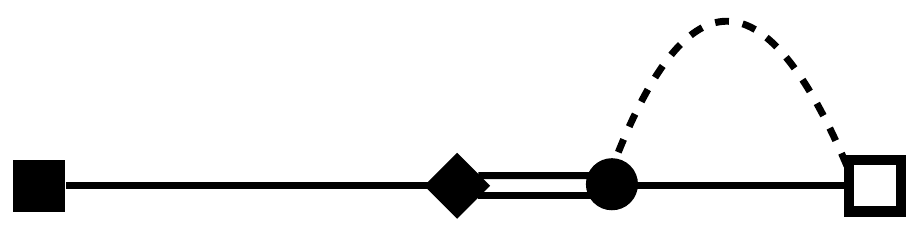}\hspace{0.5cm}\includegraphics[scale=0.35]{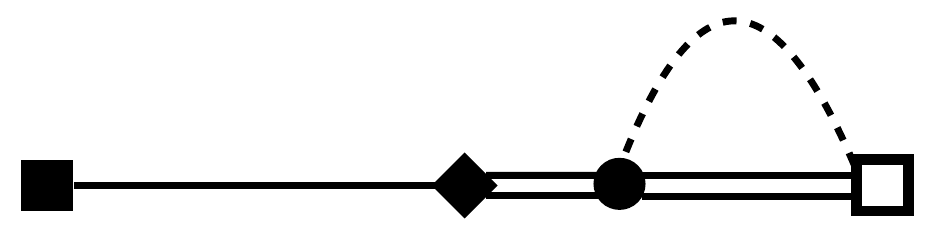}\\[0.4ex]
g)\hspace{3.3cm} h) \hspace{3.3cm} i) \hspace{3.3cm} j) \\[2ex]
\includegraphics[scale=0.35]{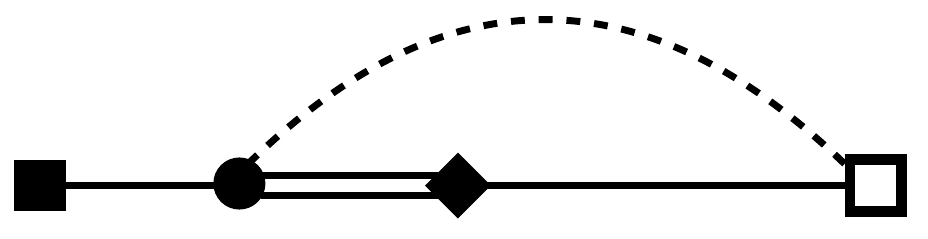}\hspace{0.5cm}\includegraphics[scale=0.35]{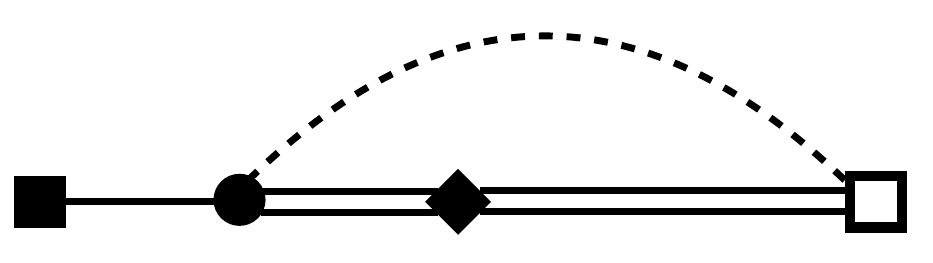}\hspace{0.5cm}\includegraphics[scale=0.35]{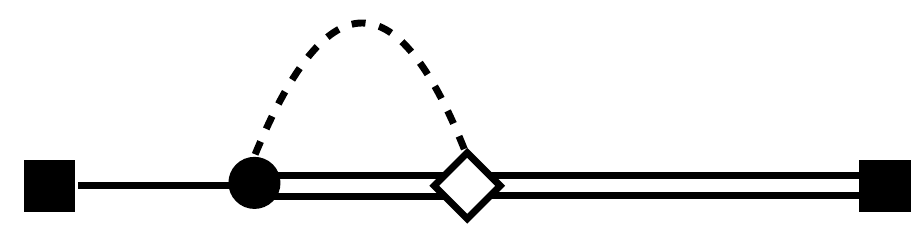}\hspace{0.5cm}\includegraphics[scale=0.35]{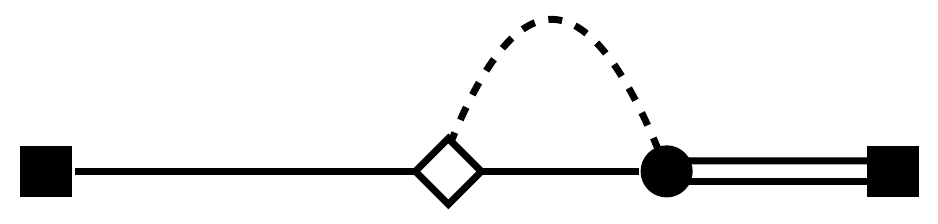}\\[0.4ex]
k)\hspace{3.3cm} l) \hspace{3.3cm} m)  \hspace{3.3cm} n)\\[0.4ex]
\includegraphics[scale=0.35]{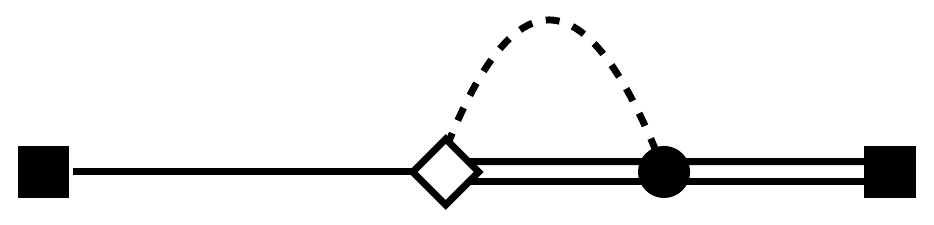}\hspace{0.5cm}\includegraphics[scale=0.35]{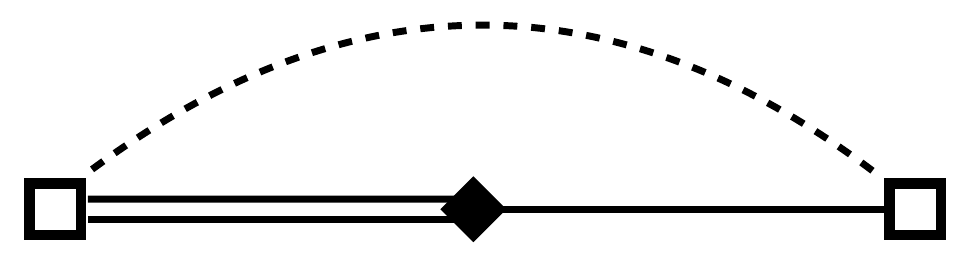}\hspace{0.5cm}\includegraphics[scale=0.35]{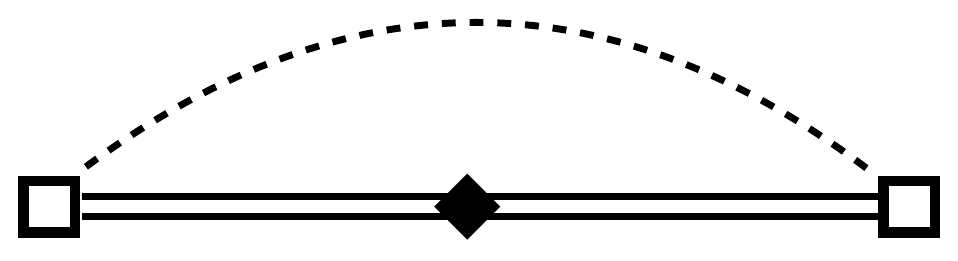}\\[0.4ex]
o)\hspace{3.3cm} p) \hspace{3.3cm} q) \\[2ex]
\caption{
Feynman diagrams contributing to the $B\pi$ contribution in 3-pt function through NLO. The open diamond represents the NLO axial vector current at operator insertion time $t'$.
}
\label{fig:diag_C3_Bpi}
\end{center}
\end{figure}

The one-loop diagrams in figure \ref{fig:diag_C3_Bpi} provide the $B\pi$ contribution in $C_3^{B\pi}$ through NLO. In analogy to \pref{DefDeltaC2} we write
\begin{equation}\label{C3specdecompGen}
C_{3}(t,t') = C^B_{3}(t,t')\left(1 + \Delta C_3^{B\pi}(t,t') \right) \,,
\end{equation}
with $\Delta C_3^{B\pi}(t,t') =  C_3^{B\pi}(t,t')/C_3^B(t,t')$. The time dependence in $\Delta C_3^{B\pi}(t,t')$ is parameterised by the coefficients $b$ and $c_{\rm 3pt}$,
\begin{equation}
\Delta C_3^{B\pi}(t,t') = \sum_{\vec{p}}\left( b(\vec{p}) \,\big[e^{-\Ep (t-t')} +  e^{-\Ep\, t'}\big] + c_{\rm 3pt}(\vec{p}) e^{-\Ep\, t}\right)\,.
\end{equation}
As in \pref{DefC2ptCoeff} it is useful to express these coefficients as a product of the universal factor $U(\vec p)$ and a reduced coefficient. We introduce capital letters $B, \, C_{\rm 3pt}$ for the latter. The diagrams in fig.\ \ref{fig:diag_C3_Bpi} yield the following results for them:
\begin{eqnarray}
b(\vec{p})&=&U(\vec{p})\,B(\vec{p})\,,\;B(\vec{p}) = \frac{8}{9} \left(g^2 + (g\tba  + \gamma) \Ep \right) \,,\label{resultNLOBandBt}\\
c_{\rm 3pt}(\vec{p})&=&U(\vec{p})\,C_{\rm 3pt}(\vec{p})\,,\;C_{\rm 3pt}(\vec{p}) = \frac{1}{9} \left(g + \tba\Ep \right)^2\,.
\end{eqnarray}
Here we have combined two  LECs of the NLO Lagrangian, \pref{NLOHMLag}, into 
\begin{equation} \label{eq:gamma}
\gamma \equiv d_3 - d_1 / 4\,.
\end{equation}

As for the 2-pt function, the LO contributions in $B,C$ are the ones proportional to $g^2$. Note that at LO the coefficient $B$ is eight times larger than $C_{\rm 3pt}$. 

The deviation $\Delta R^{B\pi}(t,t')$ for the ratio of the 3-pt and the 2-pt function is defined in eq.\ \pref{DefDeltaBpi}. Approximating $1+ \Delta C_2^{B\pi}(t)\approx [1- \Delta C_2^{B\pi}(t)]^{-1}$ we obtain $c=c_{\rm 3pt}-c_{\rm 2pt}$ and
\begin{equation}
C(\vec{p})  = C_{\rm 3pt}(\vec{p}) - C_{\rm 2pt}(\vec{p}) = -\frac{8}{9} \left(g + \tba\Ep \right)^2\,.\label{resultNLOCcomb}
\end{equation}
The functions $c=U\,C$ and $b=U\,B$ enter the midpoint and summation estimates 
\pref{e:gmid} and \pref{e:gsum} for $g_\pi$.
Note that the dominant contribution in $C$ stems from the 2-pt function in the ratio, leading to identity
\begin{eqnarray}
B^{\rm LO}  = - C^{\rm LO}\,.
\end{eqnarray}
In Refs.\ \cite{Tiburzi:2009zp,Tiburzi:2015tta,Bar:2016uoj} the two-particle $N\pi$-state contribution in the nucleon 2-pt and 3-pt function has been computed to LO in the chiral expansion. There literally the same results are found if we perform the replacement  $g\rightarrow g_A$ in our results here, with $g_A$ being the axial charge of the nucleon. 
%
\section{Estimates for the NLO LECs $\beta_1$ and $\beta_2$}
\label{sect:ExtrLECs}
%
\begin{figure}[tp]
\begin{center}
\includegraphics[scale=0.39]{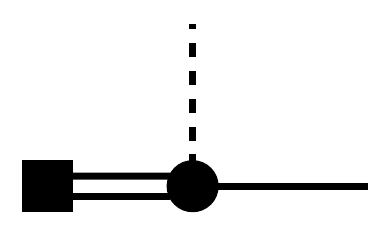}\hspace{0.5cm}\includegraphics[scale=0.39]{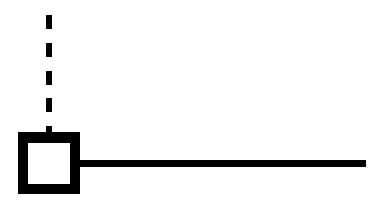}\hspace{1cm}
\includegraphics[scale=0.39]{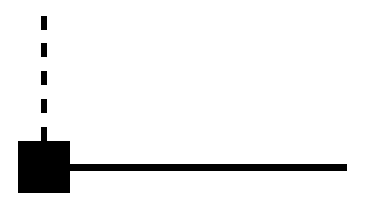}\hspace{0.5cm}\includegraphics[scale=0.39]{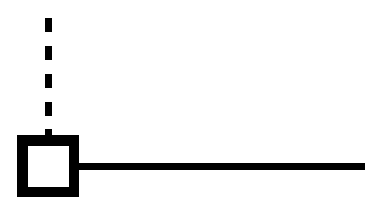}\\
a) \hspace{1.4cm} b)\hspace{2.2cm} c) \hspace{1.6cm} d)
\caption{
Tree-level Feynman diagrams contributing to the matrix elements in eq.\ \pref{e:fperp}, subfigure a) and b),  and in eq.\ \pref{e:fparr}, subfigure c) and d). The squares represent the local heavy-light vector current $V_k$ and $V_4$.
}
\label{fig:diags_VectCurFF}
\end{center}
\end{figure}

From \pref{VkExpl} we see that the low energy constants 
$\beta_i$ yield tree-level $B\pi$ couplings in the HMChPT currents. They thus contribute at tree level to the $B\to\pi$ ground state matrix elements, see fig.\ \ref{fig:diags_VectCurFF}, with the results\footnote{This part is greatly influenced by discussions with Lorenzo Barca, Julien Frison and Andreas Risch. They  made us realize that the more complicated  
proposal of \cite{Bar:2022jlw} is equivalent to the form factors which had been computed already.}
\begin{eqnarray}	
    \label{e:fperp}
	\langle \pi(\vec{p})|V_k|B(0)\rangle &=& p_k 
	\frac{\alpha g}{\sqrt{2}f E_{\pi,\vec{p}}} \left(1-\frac{\beta_1}{g}E_{\pi,\vec{p}}\right)\, + \ldots\,,
	\\
    \label{e:fparr}
	\langle \pi(\vec{p})|V_4|B(0)\rangle &=& -i 
	\frac{\alpha }{\sqrt{2}f } \left(1-{\beta_2}E_{\pi,\vec{p}}\right)\,+\ldots\,,	
\end{eqnarray}
where the ellipses contain loop contributions as well as NNLO contributions. The structure of the
analytic terms  proportional to $E_\pi$ in the above expressions 
have been given in \cite{Aubin:2007mc}, but their identifications with specific LECs only follows
from our HMChPT representation of the currents. 
The  matrix elements \pref{e:fperp} and \pref{e:fparr} have been computed in $2+1$-flavor lattice simulations of QCD in 
\cite{FermilabLattice:2015mwy,Dalgic:2006dt,Colquhoun:2022atw}. We use the results of the most recent 
computation of the KEK group~\cite{Colquhoun:2022atw}, which covered a range of $0.3\,\mathrm{GeV} \lessapprox E_\pi \lessapprox 1\, \mathrm{GeV}$ and parameterized the energy dependence 
in a form entirely compatible with the tree-level 
form.\footnote{Since the cited computations are at finite mass of the b-quark, they have a term which involves the $B^*-B$  mass splitting in their fit functions. 
Setting it to zero, the fitted energy dependence is 
exactly given by our NLO formulae up to the NNLO effects $O(E_\pi^{2})$.} We can directly identify their fit coefficients $C_E N_E$ and $D_E N_E$ with the LECs in the form
\begin{eqnarray}	
 	D_E N_E = -\beta_1/g &\;\to\;& \beta_1 = 0.14(4) \,\mathrm{GeV}^{-1}\,,\label{measbeta1}
 	\\
    C_E N_E = -\beta_2 &\;\to\;& \beta_2 = 1.2(3) \,\mathrm{GeV}^{-1}\,.\label{measbeta2}
\end{eqnarray}
Here, we used $g=0.49$ \cite{Gerardin:2021jch} and our error estimate is conservatively taken to be twice the statistical 
error of the fitted $D_E,C_E$. The reason for this caution is that the computation of the form factors~\cite{Colquhoun:2022atw} is not in the static limit and the fit form did not contain
energy-dependent discretization effects. The low energy constants of
interest describe the energy dependence and therefore are 
expected to be sensitive to possible energy dependent discretization effects. 

The above yields very useful estimates for the LECs 
of the local currents. In principle they could 
even be extracted from the experimental $B\to\pi\ell\nu$ 
decay rates, but that would result in large errors since
the rate is low at small $E_\pi$.

In practice smearing is used and one needs also $\tilde \beta_i$ of the smeared fields. For our illustrations below, we will assume that smearing 
leads to a reduction of the excited state effects and thus consider setting $\tilde \beta_i=\beta_i$ as a plausible upper bound
for the excited state effects. In the future one should determine 
$\tilde \beta_i$ directly. A rather straight forward way is to compute the ratios 
\begin{eqnarray}	
    \label{e:fperprat}
	\frac{\langle \pi(\vec{p})|\tilde V_k|B(0)\rangle}{\langle \pi(\vec{p})|V_k|B(0)\rangle}
	 &=& 
	\frac{\tilde \alpha}{\alpha}\left(1+\frac{\beta_1-\tilde\beta_1}{g}E_{\pi,\vec{p}}\right)\, + \ldots\,,
	\\
    \label{e:fparrrat}
	\frac{\langle \pi(\vec{p})|\tilde V_4|B(0)\rangle}{\langle \pi(\vec{p})|V_4|B(0)\rangle} &=&  	
\frac{\tilde \alpha}{\alpha}\left(1+ (\beta_2-\tilde\beta_2)E_{\pi,\vec{p}}\right)\, + \ldots\,,
\end{eqnarray}
together with the ratio of leading order constants, $\tilde \alpha/\alpha$. The latter is easily obtained from the 
ratio of smeared-smeared to local-smeared two-point correlation
functions or from \pref{e:fperprat} at a second energy. 
In turn, the ratios \pref{e:fperprat} and \pref{e:fparrrat} are, e.g., 
directly given by the large time behavior of  ratios of three-point functions,
$\langle \pi(t) \tilde V_\mu(t_\mathrm{v}) B(0)\rangle / \langle \pi(t) V_\mu(t_\mathrm{v}) B(0)\rangle$. Here the interpolating fields $B,\pi$
can be smeared. There are of course finite time contaminations 
in all these 2-point and 3-point functions, but given 
that one is interested in sub-leading LECs, it is expected that it is 
sufficient to just take reasonably large times in order to determine
appropriate estimates.
Once more precision is needed, the $B\pi$ contaminations can be taken into
account since they contain just the same LECs. The $B\pi$ contributions to the three-point functions needed in such a refinement will be discussed in \cite{Alexander}.

A determination of the LEC combination $\gamma=d_1-d_3/4$ requires additional effort. Expanded, the terms $O_1^{(2)}$ and $O_3^{(2)}$ involve the product of
two $B$ / $B^*$  fields and the field $
\partial_k \pi$. Therefore,  4-point correlation functions seem to be necessary in order to get a non-vanishing contribution from tree-graphs in HMChPT. We plan to study this issue in the near future. 

\section{Estimating the impact on lattice calculations}
%
\subsection{General remarks} 
\begin{figure}[tbp]
\begin{center}
$\Delta C_2^{B\pi}(t)$ \hspace{6.5cm} $\Delta C_2^{B\pi}(t)$\\
\includegraphics[scale=0.6]{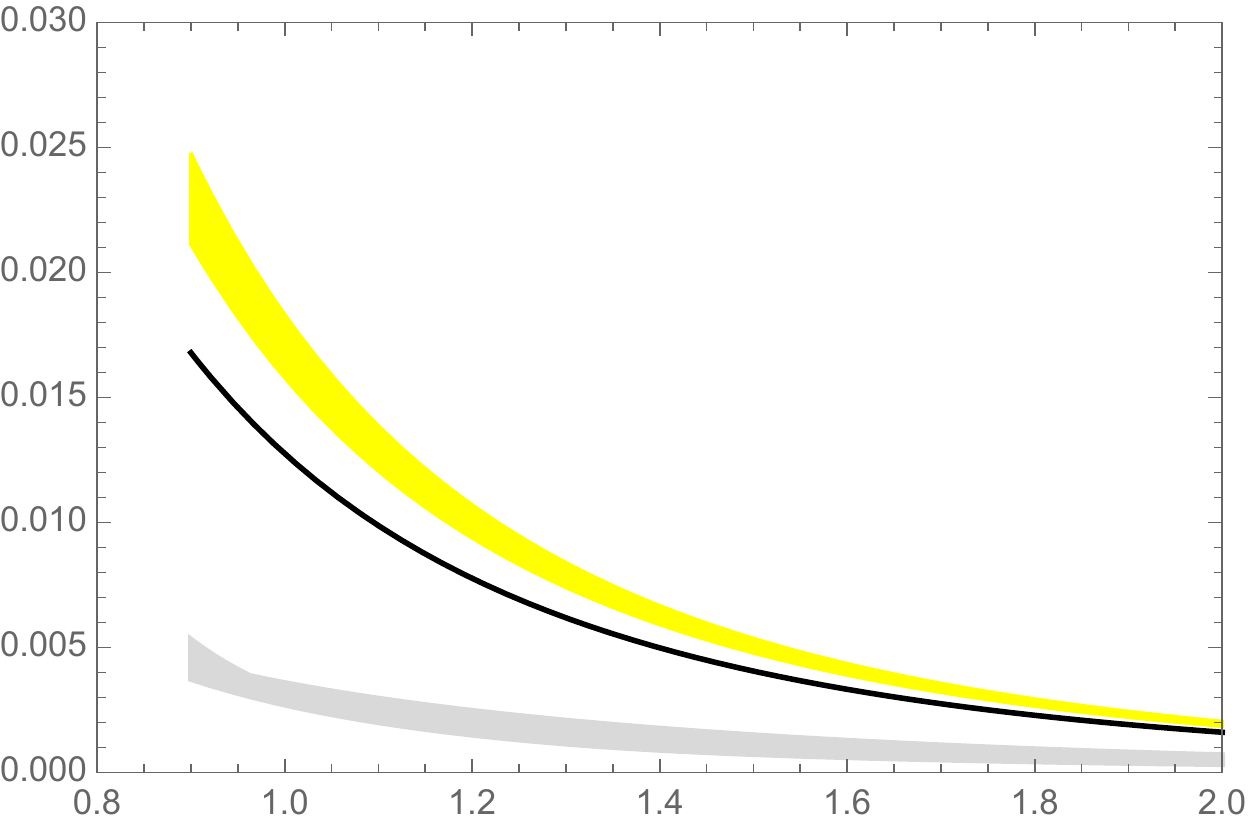}\hspace{0.7 cm}\includegraphics[scale=0.6]{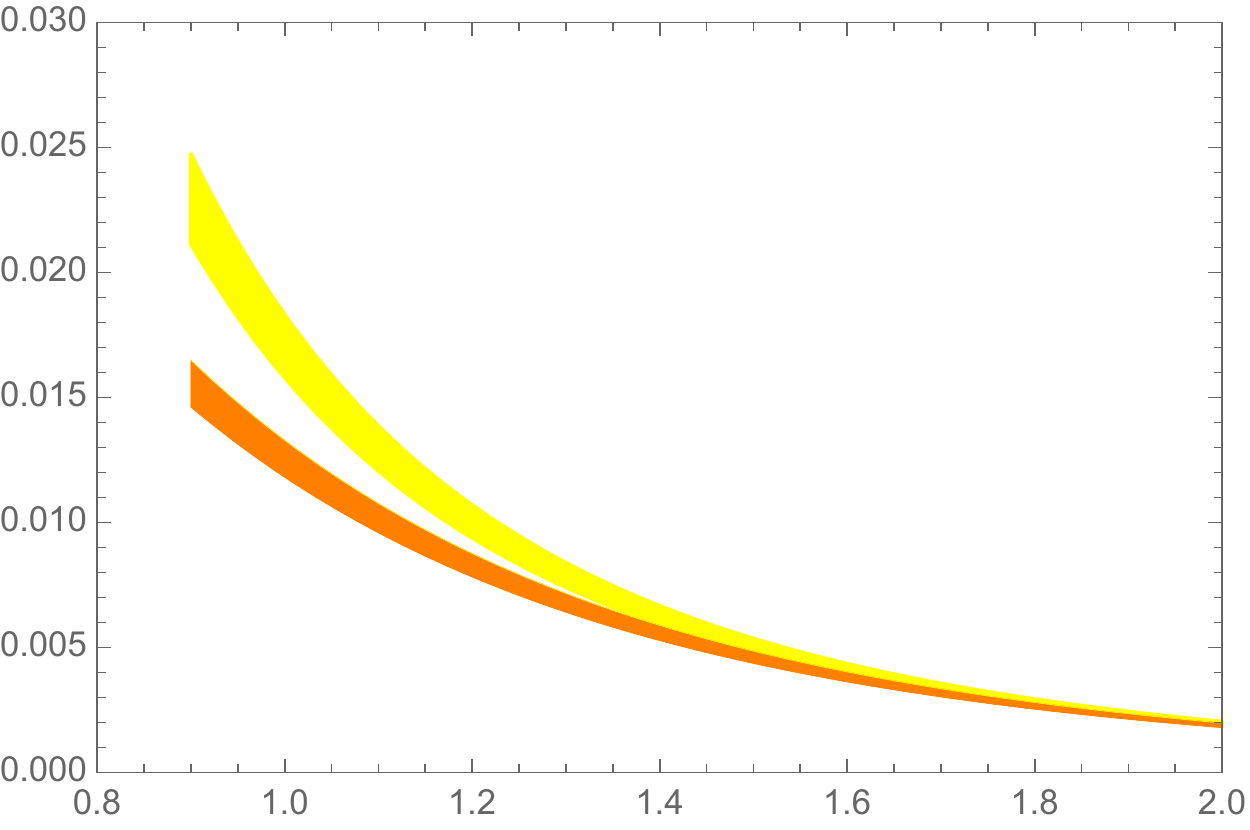}\\[2pt]
$t$/[fm]\hspace{7.5cm}$t$/[fm]\\
\caption{The deviation $\Delta C_2^{B\pi}(t)$ as a function of the source-sink separation $t$. Left panel: The solid line is the LO result, the yellow and gray band show the NLO results with $\tba \Lambda_{\chi}=0.16(5)$ and $-0.80(25)$, respectively (see main text). Right panel: The orange band shows the NLO result for $\tba \Lambda_{\chi}=0.16(5)$ and with an upper bound on the pion energy, $E\le 5 M_{\pi} = 700$ MeV. The contribution of these low-energy pion states dominate $\Delta C_2^{B\pi}(t)$ as long as $t\gtrsim 1.3$ fm.}
\label{fig:NImp_2pt_funct}
\end{center}
\end{figure}
Of practical importance is the question how big the $B\pi$ excited state contamination is at a given source-sink separation. With some reasonable assumptions our ChPT results provide an answer to this question.

Working in infinite spatial volume and at LO the ChPT results depend on two dimensionless parameters only, the ratio $f/M_{\pi}$ and the LO LEC $g$. In case of a finite spatial volume we also need to provide the size of the spatial volume by specifying $M_{\pi}L$. To the order that we are considering, it is sufficient to estimate $f$ by its phenomenological value $f\approx f_{\pi}=93$~MeV and use the lattice determination $g=0.49$~\cite{Gerardin:2021jch}.  In the following we set the pion mass to its (approximate) physical value $M_{\pi}=140$ MeV and illustrate the $B\pi$ contamination for physical point simulations.
With these values the LO ChPT results are fixed and provide predictions for the $B\pi$ contamination in the various estimators that we introduced before.

At NLO three more LECs enter: $\tba$ associated with the smeared interpolating $B$-meson fields, and the coefficients $\beta_{1}$ and $\gamma$ stemming from the heavy-light and light-light axial vector current.
All these NLO LECs have mass dimension $-1$, and this scale is expected to be set by the chiral scale $\Lambda_{\chi}$ of order 1 GeV. Since we have already introduced the scale $f_{\pi}$ we set $\Lambda_{\chi}=4\pi f_{\pi}\approx 1.2\,{\rm GeV}$, and write each LEC as a dimensionless number times $\Lambda_{\chi}^{-1}$.  This dimensionless number is expected to be of O($1$).
For the two LECs $\beta_1$ and $\beta_2$ this expectation agrees with the values determined in the last section. The values in \pref{measbeta1} and \pref{measbeta2} translate into $\beta_1=0.16(5)\Lambda_{\chi}^{-1}$ and $\beta_2=1.4(4)\Lambda_{\chi}^{-1}$.
To date no estimates for the LECs $\tba$ and $\gamma$ are available, at least to our knowledge.
As discussed in the previous section, we assume $\beta_1=\tba$. For $\gamma$  we choose to vary it between $-1.0$ and $+1.0$ in accordance with the naive dimensional analysis.

\begin{figure}[tbp]
\begin{center}
$\Delta C_2^{B\pi}(t)$\\
\includegraphics[scale=0.65]{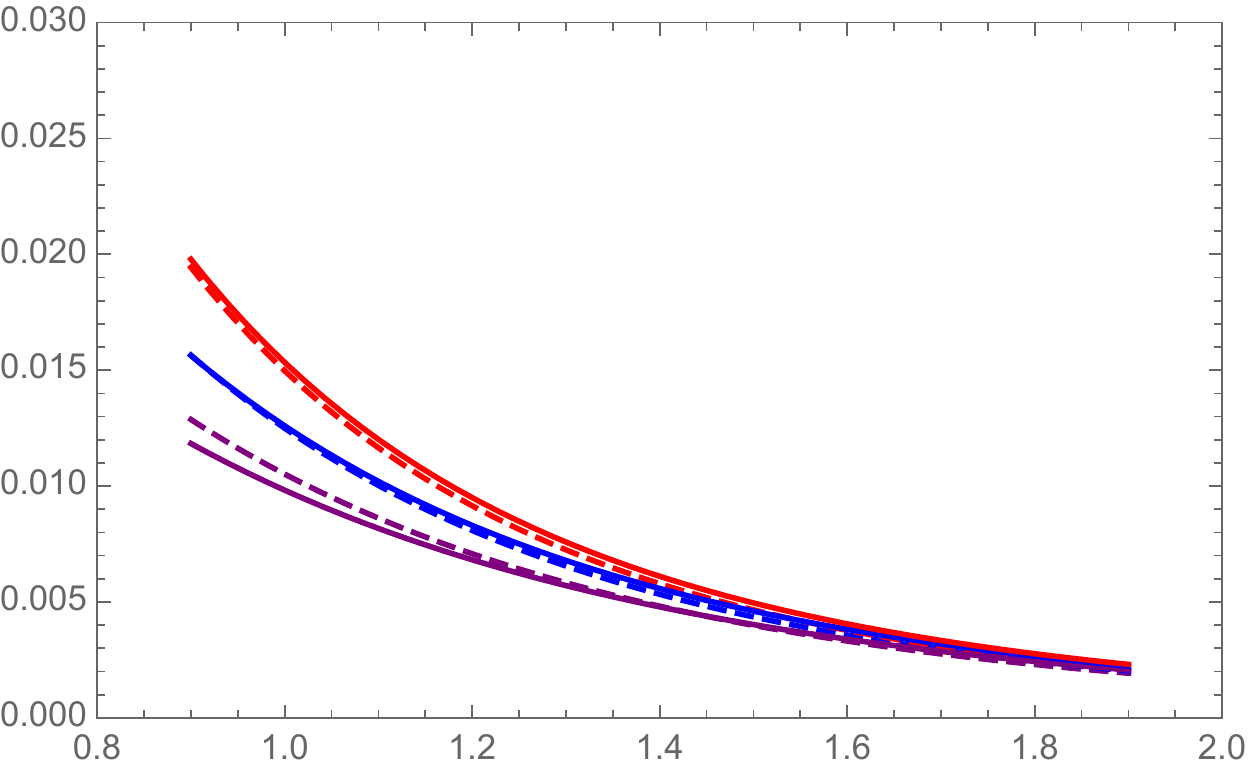}\\[2pt]
$t$/[fm]\\
\caption{The deviation $\Delta C_2^{B\pi}(t)$ as a function of the source-sink separation $t$, both for an infinite spatial volume (solid lines) and a finite volume with $M_{\pi}L=4$ (dashed lines). Results are shown for three different upper bounds on the pion energy:  $E/M_{\pi} \le 7$ (red), 5 (blue) and 4 (purple). The difference is at the sub-permille level for $t\gtrsim 1.3$ fm. 
}
\label{fig:NImp_2pt_funct_FV_Comp}
\end{center}
\end{figure}

With these assumptions about the LECs the left panel in figure \ref{fig:NImp_2pt_funct} shows the result for the deviation $\Delta C_2^{B\pi}(t)$ in the 2-pt function in case of an infinite spatial volume, see eq.\ \pref{Eq:DeltaC2_Inf_Vol}. The solid black line shows the LO result, the yellow band corresponds to the variation $\tba\Lambda_{\chi}= 0.16(5)$. 

Before reading off a value for $\Delta C_2^{B\pi}$ we need to make sure to apply the result where ChPT is expected to be reasonably applicable. Since ChPT is a low-energy effective theory the source-sink separation $t$ needs to be sufficiently large such that the deviation $\Delta C_2^{B\pi}(t)$ is dominated by low-energy pions.
The orange band in the right panel of figure \ref{fig:NImp_2pt_funct} shows the NLO results for $\Delta C_2^{B\pi}$ if we restrict the integral \pref{Eq:DeltaC2_Inf_Vol} to $E \le 5 M_{\pi} = 700\,{\rm MeV}$. The yellow band is the same as in the left panel taking into account all energies. 
At $t \approx 1.3 $ fm the two bands begin to overlap and the low-energy pions completely dominate $\Delta C_2^{B\pi}$.
For $t\simeq 0.9\,{\rm fm}$, on the other hand,  the high-energy pions with $E > 5 M_{\pi}$ contribute significantly to the $B\pi$ contamination. The figure suggests that we may expect the chiral expansion to be reasonably well behaved for $t\gtrsim 1.3\,{\rm fm}$. This is in agreement with the naive expectation that euclidean time-separations need to be at least 1 fm for pion physics to dominate the correlation functions.

Figure \ref{fig:NImp_2pt_funct_FV_Comp} shows $\Delta C_2^{B\pi}$ for three different upper bounds $E_{\rm max}$ on the pion energy: $E_{\rm max}/M_{\pi} = 7$ (solid red line), 5 (solid blue) and 4 (solid purple). $\tba\Lambda_{\chi}$ is set to the central value $0.16$, i.e.\ the blue solid line in fig.\  \ref{fig:NImp_2pt_funct_FV_Comp} is the middle of the orange band in fig.\ \ref{fig:NImp_2pt_funct}. Also shown (dashed lines) are the corresponding results in case of a finite spatial volume with $M_{\pi}L=4$. Except for the smallest upper bound and for small $t$ the difference between the FV and infinite volume results are at the sub-permille level. If we had plotted the bands stemming from the uncertainty in $\tba$ the difference would not be visible. 
Therefore, we ignore the FV corrections and discuss the infinite volume results only in the following. 

Note, however, that the FV results shown in fig.\ \ref{fig:NImp_2pt_funct_FV_Comp} are the cumulative contribution of many $B\pi$ states with different discrete pion momenta $\vec{p}_n=2\pi\vec{n}/L$. The three upper bounds $E_{\rm max}$ on the pion energy correspond to $N_{\rm max}=6, 10$ and $20$ in terms of the momentum label $N=|\vec{n}|^2$. Thus, the blue dashed line in fig.\ \ref{fig:NImp_2pt_funct_FV_Comp} is the $B\pi$ contamination caused by 
states with ten different energies. 
This number grows rapidly if $M_{\pi}L$ gets larger than 4.

Going back to fig.\ \ref{fig:NImp_2pt_funct} we can now read off a $B\pi$ excited-state contribution of about 1\% or smaller for $t\gtrsim 1.3$ fm. Applying the ChPT results to much smaller source-sink separations may not be justified and should be done with care.

Recall that we made the assumption $\tba=\ntba$, i.e.\ the LECs associated with the smeared and the local interpolating fields are the same. As discussed before we consider this as an upper limit for $\tba$, but in practice one expects an impact of smearing on the size of the $B\pi$ contamination via the value for $\tba$.  As an illustration of how big this impact can be fig.\ \ref{fig:NImp_2pt_funct} also shows the NLO result for $\tba = -5 \ntba = -0.80(25)$ (gray band). In this case the $B\pi$ contamination is drastically reduced, roughly by a factor 5 at $t\approx 1.3$ fm. 
Whether a smearing exists that can come close to $\tba = 0.80$ is an important question that needs to be studied in specific lattice simulations (see sect.~\ref{sect:ExtrLECs}). 

\begin{figure}[p]
\begin{center}
$\Delta M^{B\pi}_B(t)/ [{\rm MeV}]$\\
\includegraphics[scale=0.62]{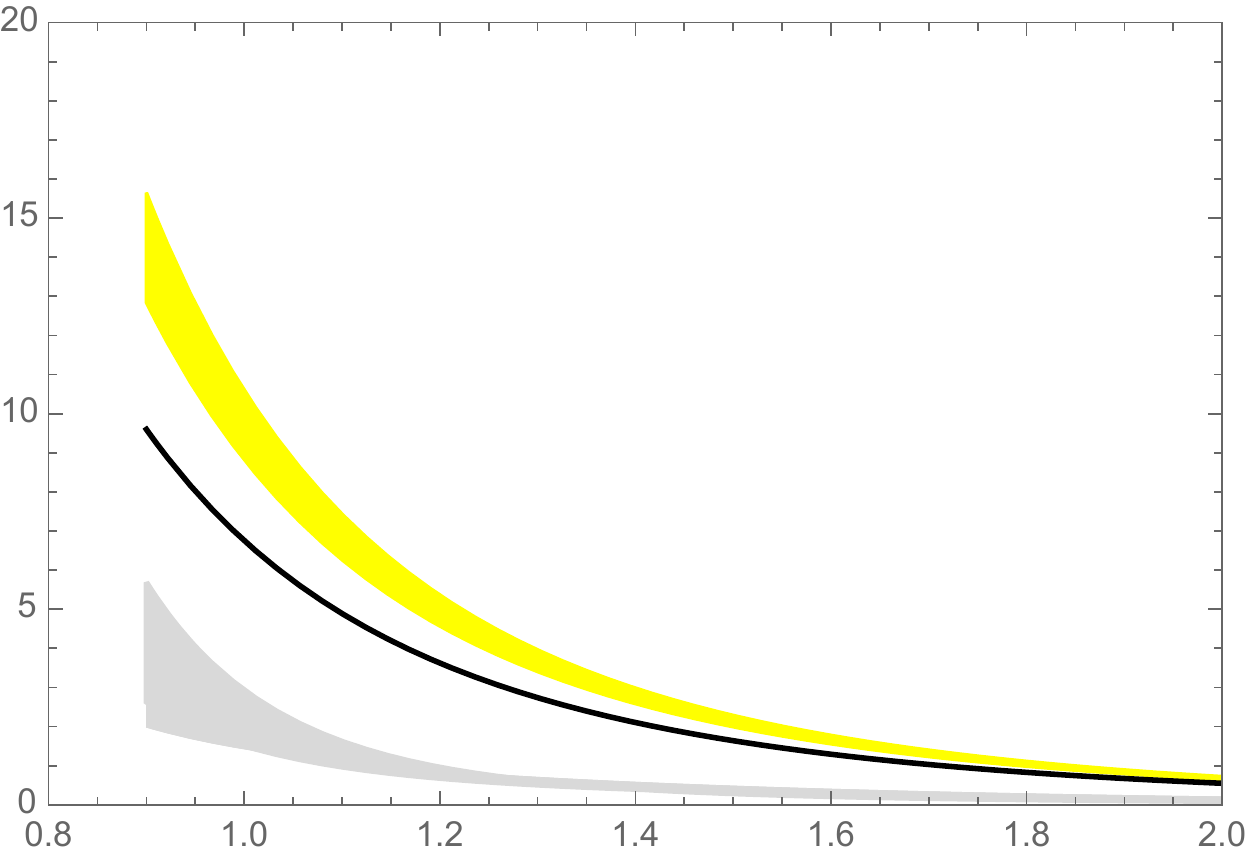}\\[2pt]
$t$/[fm]\\
\caption{The deviation $\Delta M^{B\pi}_B(t)$ as a function of the source-sink separation $t$, given in MeV. The solid line is the LO result, the yellow and gray band show the NLO results with $\tba \Lambda_{\chi}=0.16(5)$ and $-0.80(25)$, respectively.}
\label{fig:NImp_EffMass}
\end{center}
\begin{center}
$\Dfhateff(t)$\\
\includegraphics[scale=0.62]{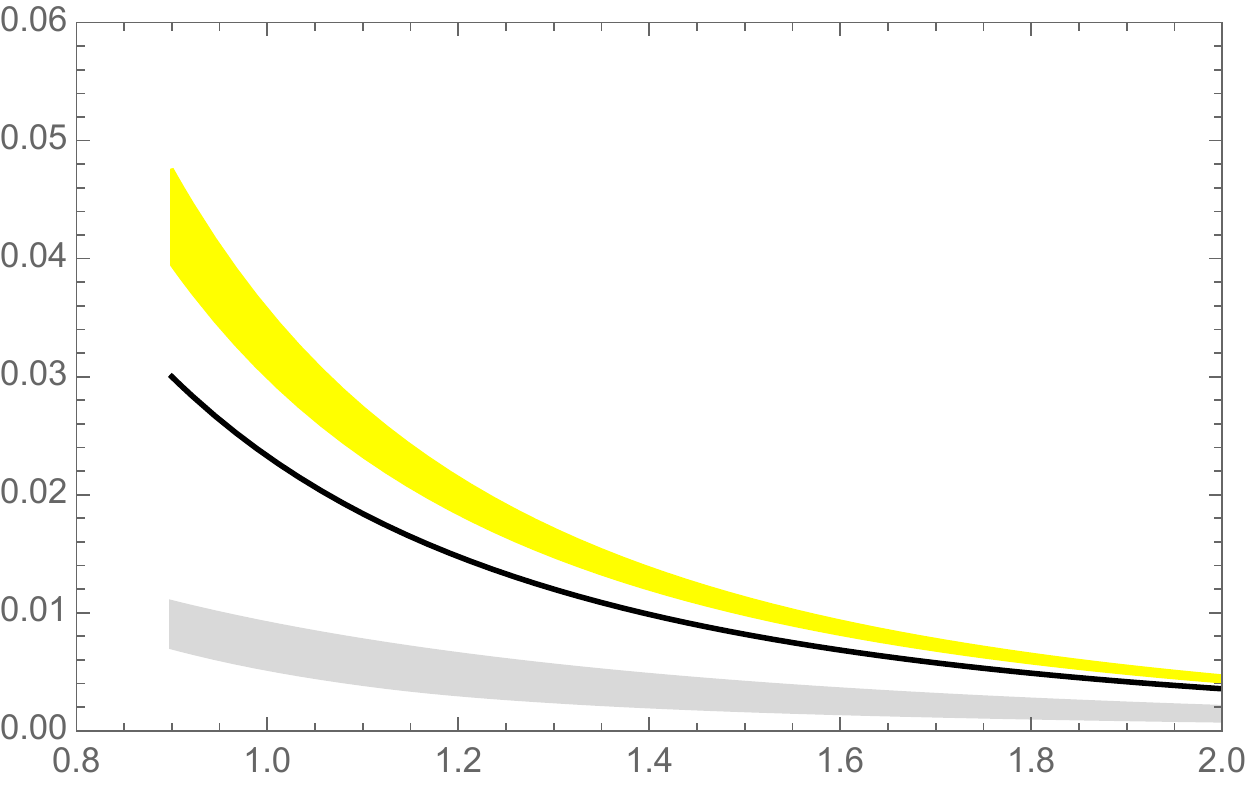}\\[2pt]
$t$/[fm]\\
\caption{The deviation $\Dfhateff(t)$ as a function of the source-sink separation $t$. The solid line is the LO result, the yellow band shows the NLO result with both NLO LECs $\tba \Lambda_{\chi}$ and $\ntba \Lambda_{\chi}$  being varied between $0.11$ and $0.21$.
The gray band shows the NLO result for $\ntba=0.16(5)\Lambda_{\chi}^{-1}$ and $\tba= -0.8(25)\Lambda_{\chi}^{-1}$ as an illustration for the theoretically possible impact of smearing.
}
\label{fig:NImp_fhatSquared}
\end{center}
\end{figure}

\subsection{Impact on the effective mass and effective decay constant}

With the preparations of the last subsection we turn to the $B\pi$ contribution in the two observables based on the 2-pt function, the effective mass and effective decay constant.

Figure \ref{fig:NImp_EffMass} shows the dimensionful deviation $\DMeff(t)$ in \pref{Res_DMeff} given in MeV. As before the solid line shows the LO result, and the yellow band the NLO result with   the  LEC $\tba$ varied as discussed above. At $t=1.3$ fm the NLO result is about 4 MeV.

Figure \ref{fig:NImp_fhatSquared} displays the deviation $\Dfhateff(t)$ in \pref{Res_Deltafhat} for the effective decay constant. Here the deviation is roughly 1.6\% at $t=1.3$ fm. 

As before, the figures also show the NLO results for $\tba=-5 \ntba$ (gray bands), the value that leads to a substantial reduction of the $B\pi$ contamination in the 2pt function. Expectedly, this reduction also leads to a much smaller $B\pi$ contamination in the effective mass and effective decay constant.

\subsection{Impact on the 3-pt function and the $B^*B\pi$ coupling}

\begin{figure}[tbp]
\begin{center}
$\Dgsum(t)$ \hspace{6.5cm} $\Dgmid(t_{\rm mid})$\\
\includegraphics[scale=0.6]{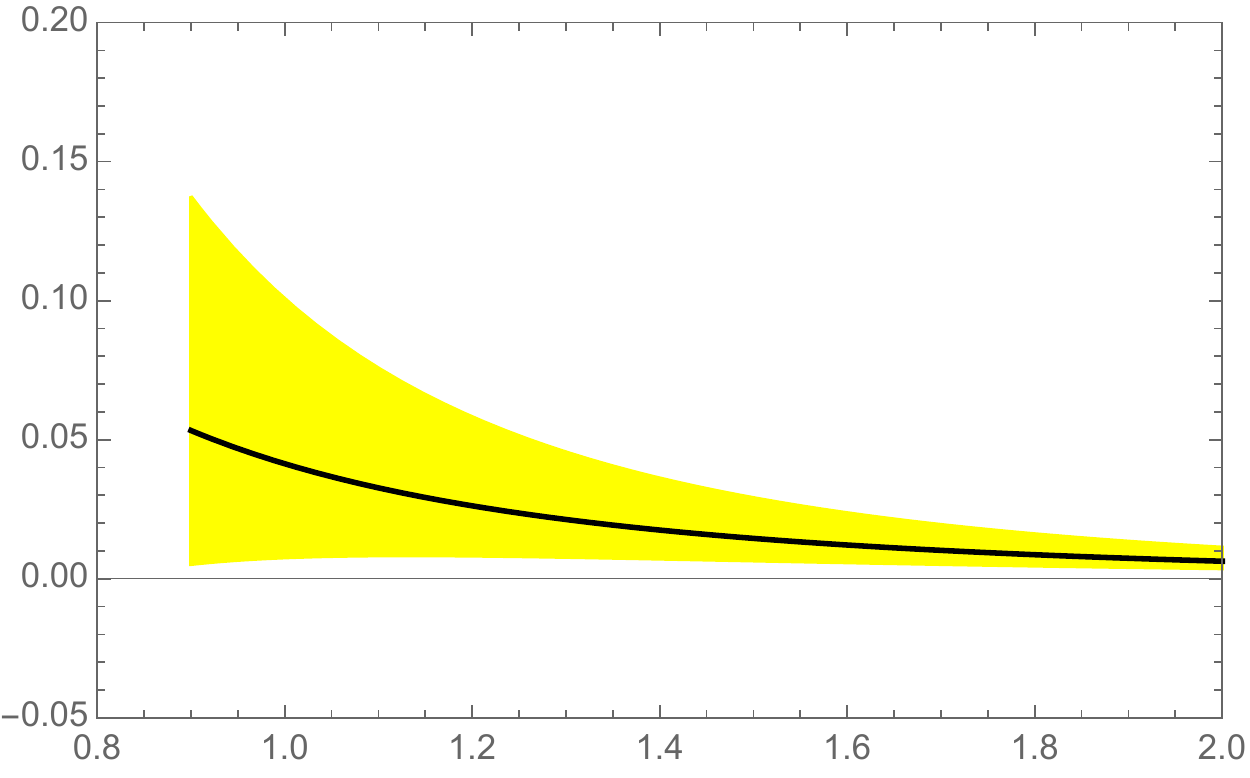}\hspace{0.7 cm}\includegraphics[scale=0.6]{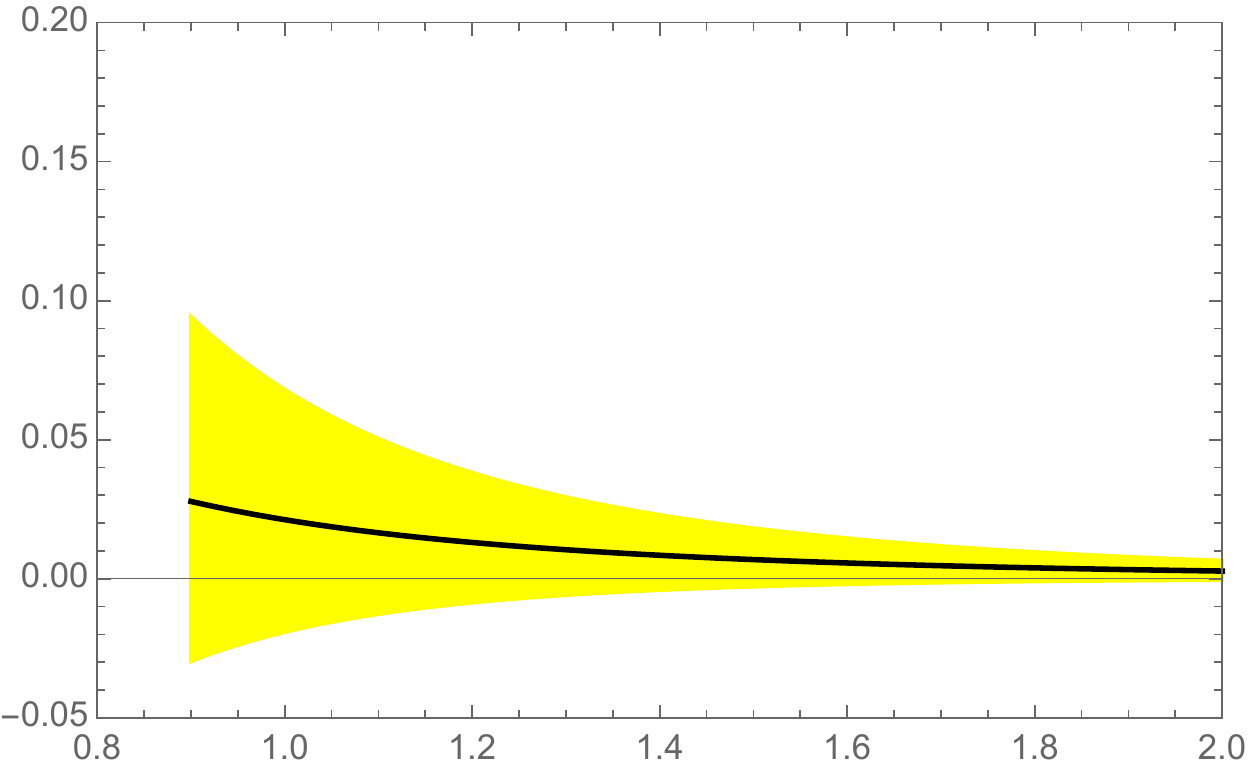}\\[2pt]
$t$/[fm]\hspace{7.5cm}$t_{\rm mid}$/[fm]\\
\caption{The deviations $\Delta g^{{\rm sum},B\pi}_{\pi}$ (left panel) and $\Delta g^{{\rm mid},B\pi}_{\pi}$ (right panel). 
The solid lines are the LO results, the yellow band show the NLO results with  $0.11 \le \tba \Lambda_{\chi}\le 0.21$ and $\gamma \Lambda_{\chi}$  being varied between 1.0 and $-1.0$. Note that $\Delta g^{{\rm mid},B\pi}_{\pi}$ is plotted as a function of $t_{\rm mid}=t/2$.  }
\label{fig:NImp_gsum}
\end{center}
\end{figure}

The left panel of figure \ref{fig:NImp_gsum} shows the deviation $\Dgsum(t)$ as a function of $t$. As before the solid line shows the LO result, the yellow band is the NLO result with the NLO coefficients being varied within our chosen bounds. Recall that $\Dgsum$ involves two NLO LECs, $\tba$ and $\gamma$, the latter being associated with the light axial vector current. 
At $t=1.3$ fm, $\Dgsum$ varies roughly between 1\% to 5\%. This spread is dominated by the lack of knowledge of the unknown LEC $\gamma$.

For illustrative purposes only we show the deviation $\Dgmid$ for the mid-point estimate in the right panel of figure \ref{fig:NImp_gsum}. 
For an appropriate comparison of the two figures we have plotted  $\Dgmid$ as a function of $t_{\rm mid}\equiv t/2$. This way the deviations are roughly of the same size, as expected. Note, however, that values $t/2\gtrsim1$ fm are hard to reach in practice. 

%
\section{Discussion}
%
We have started from the HMChPT expansion at zero velocity of the heavy quark. Because some elements -- 
the effective currents and,  in particular, the smeared interpolating fields -- had not been constructed systematically at NLO, we filled this gap. 
At NLO, there are
LECs $d_k\,,\; k=1\ldots 7$ in the Lagrangian.
Two of them find their way, following the standard construction, into the light-light axial vector current. 
One particular combination of these two, denoted by $\gamma$, enters our prediction for the estimators of
the $B^\star B \pi$ coupling. It is presently unknown and we have used power counting in the form 
$|\gamma|\leq 1/\Lambda_\chi = 1/ (4\pi f)$ to  illustrate
its effect. Due to the symmetries of HQET,
the local effective heavy-light bilinears 
are given by only two NLO LECs,
$\beta_i, \;i=1,2$. Their values were estimated from $B\to \pi$ form-factors.

The size of the $B\pi$ excited-state contamination is rather modest 
in the considered estimators: typically at the few-percent level at distances $t=1.3$~fm. 
For the two-point functions of the local heavy-light
currents, the predictions appear to be rather reliable for, say, $t\approx 1.3$~fm and larger. Our arguments for reliability are standard: i) the $B\pi$ contributions 
are significantly smaller than the leading one-particle 
ones, ii) the NLO contribution is only a small part of
the leading order in the chiral expansion and iii) the 
uncertainties of the LECs are small enough to make such statements. In fact, it appears best to
use the NLO contribution as the overall uncertainties of our predictions.

Concerning the use of smeared interpolating fields,
these require LECs $\tilde \beta_i(r_\mathrm{sm})$
which depend on the smearing radius (apart from the smearing type itself). 
It is very plausible that smearing will reduce the excited state 
contaminations, but how much is not  understood. We therefore recommend the results for local interpolating fields as upper bounds for
the effects of $B\pi$ states as long as the $\tilde \beta_i(r_\mathrm{sm})$ are not known. 

The next important step is to determine the $\tilde \beta_i(r_\mathrm{sm})$ as we described in section \ref{sect:ExtrLECs}. It is very interesting to study how a concrete smearing procedure affects their values and, consequently, the size of the $B\pi$ contamination. The effect can be quite dramatic, as we have illustrated in figs.\ \ref{fig:NImp_EffMass} and \ref{fig:NImp_fhatSquared}.

We allow ourselves to 
speculate on $\gamma$. Throughout the history of numerical determinations of the $B^\star B \pi$ coupling \cite{Bernardoni:2014kla,deDivitiis:1998kj,Ohki:2008py,Becirevic:2009yb,Detmold:2012ge,Flynn:2013kwa}, excited state contaminations have been found to be very small in this channel. This suggests that $\gamma$ is at the border of our considered range
and compensates (much of) the leading order effects. A  caveat to this statement is that the numerical computations have typically {\em not} been done in the range where we trust HMChPT but at smaller $t$. We thus need to assume that the flat behavior seen in the numerical determinations continues to larger $t$.

In principle it is possible to go to the NNLO chiral order, however, it is doubtful whether this is useful in practice. Quite a few additional LECs will enter the results, and we expect most of them to be not easily accessible.

More interesting is the computation of the $B\pi$ contamination in other $B$-physics observables, for instance the form factors relevant for the semi-leptonic decay $B\rightarrow\pi$ and the determination of the CKM matrix element $V_{ub}$. A first study shows \cite{Bar:2022ksk} that the LECs $\tilde \beta_i$ also enter the $B\pi$ contamination in the effective form factors. This renders the lattice computation of these LECs even more desirable.
\vspace{0.5cm}

{\noindent \large\bf Acknowledgements}\\

AB is funded by the {\it Deutsche Forschungsgemeinschaft}  (German Research Foundation, DFG) - project number 417533893/ GRK2575 “Rethinking Quantum Field Theory”.
We thank Lorenzo Barca, Julien Frison and Andreas Risch  for very useful discussions on the extraction of the
low energy constant $\beta_1$.

\newpage

{\large\bf Appendix}

\begin{appendix}

\section{Feynman rules}\label{app:Feynman_rules}

The kinetic part of \pref{LOHMLag} consists of the individual kinetic parts for the pseudoscalar and vector meson fields with the known propagators \cite{Bernardoni:2009sx}. All of our calculations are done in the RF of the heavy meson, so we quote the propagators for this special case only, 
\begin{eqnarray}
\langle P_r(x) P_{{r'}}^{\ast}(y) \rangle & =&\delta_{rr'}G_P(x,y)\,,\label{PropP}\\
\langle P_{k,r}(x) P_{k',r'}^{\ast}(y) \rangle & =&\delta_{kk'} \delta_{rr'}G_P(x,y)\,.\label{PropVM}
\end{eqnarray}
Here $r,r'=1,2$ denote the SU(2) flavor indices, and $k,k'=1,2,3$ stand for the spin degrees of freedom for the vector meson propagator. The remaining part is given by
\begin{equation}
 G_P(x,y) \,=\, \frac{1}{2}\theta(x_4 - y_4) \delta(\vec{x}-\vec{y})\,.
\end{equation}
In the static limit the heavy mesons propagate forward in time only.

The pion propagator is obtained from the Lagrangian in \pref{L2PureMesonic} by expanding $u_{\mu}$ and $\chi_+$ to quadratic order in the pion fields. In a finite spatial volume $L^3$ it is given by
\begin{equation}\label{scalprop}
G^{ab}_{\pi}(x,y)=   \delta^{ab}L^{-3}\sum_{\vec{p}} \frac{1}{2 E_{\pi,\vec{p}}} e^{i\vec{p}(\vec{x}-\vec{y})} e^{-E_{\pi,\vec{p}} |x_4 - y_4|}\,,\quad a,b\,=\,1,2,3\,,
\end{equation} 
with 
pion energy $E_{\pi,\vec{p}} =\sqrt{\vec{p}^2 +M_{\pi}^2}$. Throughout we assume isospin symmetry so all three pions are mass degenerate.

For the $B\pi$ contamination in the correlation functions considered in this paper we need the interaction vertices that couple two heavy mesons to one pion. Vertices with more than one pion field contribute either to multi-particle state contributions with more than one pion, or to loop corrections in the $B\pi$ contamination. The latter is of higher order in the chiral expansion than considered in this paper.

The vertices stemming from the covariant derivative in ${\cal L}^{(1)}_{\rm HM}$ as well as all terms in ${\cal L}^{(2)}_{\rm HM}$ involve an equal number of pion fields, thus they can be ignored for our purposes. The relevant vertices stem from the term proportional to $g$ in ${\cal L}^{(1)}_{\rm HM}$. In the RF we explicitly find
\begin{eqnarray}\label{Vertex1PiRestframe}
{\cal L}^{\vec{v}=0}_{{\rm int},1 \pi} &=& \frac{2ig}{f}\Big(P_{k,r}\partial_{k}\pi^aT^a_{rs} P^{\ast}_{s}  -P_r\partial_{k}\pi^aT^a_{rs} P_{k,s}^{\ast}+\epsilon_{klm}P_{k,r} \partial_{l}\pi^aT^a_{rs} P_{m,s}^{\ast}\Big)\,.
\end{eqnarray}
Recall our convention $T^a=\sigma^a/2$ for the generators of the flavour group SU(2).
Diagramatically these interaction vertices are depicted in fig.\ \ref{fig:int_vertices}.

The general chiral expressions for the interpolating fields have been derived in detail in section \ref{ssect:smearedInterpolatingFields}. For our application we need the chiral expressions for the smeared interpolating fields for the heavy pseudoscalar and vector meson, $\tilde{P} = \langle \tilde{O}_V\gamma_{5}\rangle$ and $\tilde{V}_{k} = \langle \tilde{O}_V\gamma_{k} \rangle$. Expanding the individual terms in \pref{defOtildeV} in pion fields and dropping all terms with more than one pion field we find at LO
\bea
\tilde{P}_r^{(0)}&=& \tilde{\alpha} P_r\,,\label{FRLOPtilde}\\
\tilde{V}_{k,r}^{(0)}&=& \tilde{\alpha} P_{k,r} \,,
\eea
where we made use of eq.~\eqref{RedefTLECs}.
For the NLO part with chiral dimension 1 we first make use of the EOM, thus dropping the $O^{(1)}_{3,V}$ term in table \ref{table1}. In addition, we do not need to consider the term with $j=2$, since it involves at least two pion fields. Based on the remaining term we obtain the following expressions,
\bea
\tilde{P}_r^{(1)} &=&  i\frac{\tilde{\beta}_1 \tilde{\alpha}}{f}P_{k,s} \partial_k\pi^a T^a_{sr}\,,\label{FRNLOPtilde}\\
\tilde{V}_{k,r}^{(1)}&=& - i\frac{\tilde{\beta}_1 \tilde{\alpha}}{f} \Big(P_s\partial_k \pi^a - \epsilon_{klm}P_{l,s} \partial_m \pi^a \Big)T^a_{sr}  \,.\label{VkExpdim1}
\eea
Here too we have renamed the LEC entering at this order using eq.~\eqref{RedefTLECs}.
For our application the interpolating fields involve one LO and one NLO LEC.
Diagrammatically the vertices for the interpolating fields are depicted in fig.\ \ref{fig:IntField_vertices}.

For the effective decay constant \pref{DefFhatestimatorLS} we introduced the 2-pt function $C^{LS}_2(t)$ with the local current $A_4$ at the sink and a smeared interpolator at the source. The local bilinears satisfy $P_r = -A_{4,r}$, thus with \pref{FRNLOPtilde} we immediately obtain 
\bea
{A}_{4,r}^{(0)}&=& -{\alpha} P_r\,,\\
A_{4,r}^{(1)}&=&-  i\frac{{\beta}_1 \alpha}{f}P_{k,s} \partial_k\pi^a T^a_{sr} \,,
\eea
with the analogous short hand notations for the LECs associated with the local currents. 

The local field $V_4$ needed for the computation of $h_\parallel$ can be obtained by the special chiral transformation~\eqref{SpecChirTrafoCurrents} which in the effective theory replaces $u_+$ by $u_-$ and vice versa. Applying this transformation to $A_4$ in equation~\eqref{A4Expl} and expanding to linear order in the pion field leads to
\begin{align}
V_{4,r}^{(0)} &=   -i \frac{\alpha}{f} P_s \pi^a T^a_{sr}\,,\\ 
V_{4,r}^{(1)} &= i \frac{\beta_2 \alpha}{f} P_s \partial_4 \pi^a T^a_{sr}\,.   
\end{align}

Finally, for the 3pt function defined in \pref{DefC3} we need the chiral expression for the light axial vector current, which reads 
\begin{eqnarray}
A^{a}_{\mu} & = & \overline{q}_r\gamma_{\mu}\gamma_5 T^a_{rs} q_s
\end{eqnarray}
on the quark level. As already mentioned in section \ref{ssect:chiralLag}, the expression for $A^{a}_{\mu}$ in the chiral effective theory is most conveniently obtained by taking the derivative of the effective chiral action with respect to the external source field $a_{\mu}^a(x)$, cf.\ \pref{Def:lightAmu}.
At LO the source field enters the effective action via $u_{\mu}$ and the chiral connection $\Gamma_{\mu}$, see eqs.\ \pref{Def:umu} and \pref{Def:Gammamu}. 
Based on ${\cal L}^{(1)}_{\rm HM}$ we obtain the result 
\begin{eqnarray}
A^{a,(1)}_{k} &=& - 2i g \Big(P_r T^a_{rs} {P}^{\ast}_{k,s} -  P_{k,r}T^a_{rs}{P}_s^{\ast} + \epsilon_{klm}P_{l,r}T^a_{rs}P^{\ast}_{m,s} \Big) \,,\label{AxialVectEuclK1}\\
A^{a,(1)}_{4} &=& +\frac{2i}{f}\epsilon^{abc}\pi^b \Big(P_r T^c_{rs} P_{s}^{\ast} + P_{k,r}T^c_{rs}{P}^{\ast}_{k,s}\Big) \,,\label{AxialVectEucl41}
\end{eqnarray}
for the current in the rest frame. 
The purely mesonic part ${\cal L}^{(2)}_{\pi}$, contributes the familiar LO expression
\begin{equation}\label{AmumesLO}
A^{a,(2)}_{\mu,{\rm mes}} \,=\, f \partial_{\mu} \pi^a\,.
\end{equation}
In the definition of the 3pt function~\pref{DefC3}, which is needed for the computation of $g$, the axial current is projected to zero momentum and the purely mesonic part therefore does not contribute. The expressions in \pref{AxialVectEuclK1} and \pref{AmumesLO} are graphically depicted in the top row of figure \ref{fig:axial_vect_vertices}.

The chiral Lagrangian ${\cal L}^{(2)}_{\rm HM}$ contains the seven terms given in table \ref{table:NLOtermsLag}.
In addition to the definitions already given in section \ref{sect:HMChPT} two more elements enter, the Clifford algebra element $\sigma_{\mu\nu} = i[\gamma_{\mu},\gamma_{\nu}]/2$ and the source field dependent combination
\begin{eqnarray}
f_{+,{\mu\nu}} &=& u F_{R,{\mu\nu}} u^{\dagger} + u^{\dagger} F_{L,{\mu\nu}} u\,
\end{eqnarray}
with the two field strength tensors
\begin{eqnarray}
F_{R,{\mu\nu}} &=&\partial_{\mu} r_{\nu} - \partial_{\nu}r_{\mu} - i[r_{\mu},r_{\nu}]\,,\\
F_{L,{\mu\nu}} &=&\partial_{\mu} l_{\nu} - \partial_{\nu}l_{\mu} - i[l_{\mu},l_{\nu}]\,.
\end{eqnarray}
With these conventions we obtain the expression
\begin{eqnarray}
A^{a,(2)}_{k} &=& - d_1 \frac{\partial_k\pi^a}{f} \left(P_r P^{\ast}_r + P_{l,r} P^{\ast}_{l,r}\right)
\nn\\
& & + 2 i d_3 \,\epsilon^{abc}\frac{\partial_l\pi^b}{f} \Big( P_{k,r} T^c_{rs}P^{\ast}_{l,s} - P_{l,r}T^c_{rs} P^{\ast}_{k,s}  +\epsilon_{klm}[ P_r T^c_{rs}P^{\ast}_{m,s} -P_mT^c_{rs}P^{\ast}_{s} ]\Big)\,.\label{AxialVectEuclK2}
\end{eqnarray}
for the spatial components of the current in the RF. Here we have dropped the contribution proportional to $d_4$ since it is is a total derivative which does not contribute to the 3-pt function in \pref{DefC3} we are interested in. The two LECs $d_1$ and $d_3$ enter in the combination $\gamma$ defined in eq.~\pref{eq:gamma}. The time-like component $A^{a,(2)}_{4}$ is not needed in this paper. In terms of vertices in Feynman diagrams the expression \pref{AxialVectEuclK2} corresponds to the lower row of figure  \ref{fig:axial_vect_vertices}.

\section{Analytic results for the infinite volume limit}\label{app:AnalyticIVresults}
%
\subsection{Preliminary remarks} 
%
In section \ref{ssect:Results2ptfunction}, eq.\ \pref{Eq:DeltaC2_Inf_Vol} we derived the $B\pi$ excited-state contamination in the $B$-meson 2pt function for an infinite spatial volume,
\begin{equation}\label{APPEq:DeltaC2_Inf_Vol}
\Delta C_2^{SS,B\pi}(t) \,=\, \frac{3}{16\pi^2f^2} \int_{M_{\pi}}^\infty d E\, \,\frac{p^3}{E^2}   \left( g +\tba E \right)^2 e^{-E\, t}\,.
\end{equation}
The integrand is bounded and well-behaved, so the integral is easily integrated numerically using standard software packages,  MATHEMATICA \cite{Mathematica} for instance. However, as we will show in this appendix,  this and related integrals can also be expressed in terms of well-documented (modified) Bessel- and Struvefunctions \cite{NIST:Handbook,NIST:DLMF}. 

\subsection{Decomposition}
The integral in \pref{APPEq:DeltaC2_Inf_Vol} is of mass dimension two. Performing the substitution $E\rightarrow x= E/M_{\pi}$ it can be expressed as $M_{\pi}^2$ times a dimensionless integral,
\begin{eqnarray}
\Delta C^{SS,B\pi}_2(t) & = &  \frac{3}{16\pi^2}\frac{M_{\pi}^2}{f^2} H(z)\,,\\
H(z) &=& \int_1^{\infty} dx\, \frac{(x^2-1)^{3/2}}{x^2}  C_{\rm 2pt}(x) e^{-z x}\,,
\end{eqnarray}
where the function $C_{\rm 2pt}(x)$ is the coefficient in eq.\ \pref{resC2pt}, but expressed in terms of dimensionless quantities,
\begin{eqnarray} \label{APPresC2pt}
C_{\rm 2pt}(x) =
\left(g +\tilde{u}x   \right)^2\,,
\end{eqnarray}
with the short hand notation
\begin{equation}\label{Defzandu}
z \equiv M_{\pi} t\,,\qquad \tilde{u}\equiv \tilde{\beta}_1 M_{\pi}.
\end{equation}
Expanding $C_{\rm 2pt}(x) $ the integral $H(z)$ can be written as a sum of three basic integrals,
\begin{eqnarray}\label{resultH}
H(z) &=& g^2 J_0(z) + 2g\tilde{u} J_1(z) + \tilde{u}^2 J_2(z)\,,
\end{eqnarray}
with the definition
\begin{equation}\label{defJk}
J_k(z) \,=\,  \int_1^{\infty} dx\, \frac{(x^2-1)^{3/2}}{x^2} x^k e^{-z x}\,.
\end{equation}
In terms of these integrals the $B\pi$ contamination in the 2-pt function reads
\begin{eqnarray}\label{APPresultSS}
\Delta C_2^{SS,B\pi}(t) &=& \frac{3}{16\pi^2}\frac{M_{\pi}^2}{f^2} \left[g^2 J_0(z) + 2g\tilde{u} J_1(z) + \tilde{u}^2 J_2(z)\right]\,.
\end{eqnarray}
Recall that this result holds for the case with smeared interpolating fields at both source and sink, as can be infered from the quadratic dependence on $\tilde{\beta}_1$ via $\tilde{u}$.  As discussed in section \ref{ssect:Results2ptfunction}, for $\Delta C_2^{LS,B\pi}$
with a one local and one smeared interpolating field we need to perform the substitution $(g +\tilde{u}x)^2 \rightarrow(g +\tilde{u}x)(g +ux)$ and obtain 
\begin{eqnarray}\label{APPresultLS}
\Delta C_2^{LS,B\pi}(t) &=& \frac{3}{16\pi^2}\frac{M_{\pi}^2}{f^2} \left[g^2 J_0(z) + g(\tilde{u}+u)  J_1(z) + \tilde{u}u J_2(z)\right]\,,
\end{eqnarray}
with ${u}\equiv {\beta}_1 M_{\pi}$ stemming from the local interpolating field at either source or sink.

The infinite volume results for other quantities are derived the same way. For example, 
the $B\pi$ contamination in the effective $B$-meson mass, given in eq.\ \pref{Res_DMeff}, is found as
\begin{eqnarray}\label{APPresultMBeff}
\frac{\Delta M_B^{\rm eff} (t)}{M_{\pi}}  & = &  \frac{3}{16\pi^2}\frac{M_{\pi}^2}{f^2} \left[  g^2 J_1(z) + 2g\tilde{u} J_2(z) + \tilde{u}^2 J_3(z)  \right]\,.
\end{eqnarray}
The result \pref{Res_Deltafhat} for the $B\pi$ contamination $\Delta \hat{f}^{B\pi}$ in the effective decay constant reads
\begin{eqnarray}
\Delta \hat{f}^{B\pi}(t) &=& \Delta C_2^{LS}(z) - \frac{1}{2} \left(\Delta C_2^{SS}(z) - z \frac{\Delta M_B^{\rm eff} (z)}{M_{\pi}} \right)\,,
\end{eqnarray}
and with the results \pref{APPresultSS} - \pref{APPresultMBeff} we obtain the desired decomposition in terms of the basic integrals $J_k$. 

Finally, the infinite volume limit result for the $B\pi$ contamination in the summation estimate for the $B\pi$ coupling is obtained following the same steps. Taking the infinite volume limit of \pref{gPisumest} and performing again the substitution $E\rightarrow x= E/M_{\pi}$ we can establish the result
\begin{eqnarray}
\Delta g^{{\rm sum},B\pi}(t)& = &  \frac{3}{16\pi^2}\frac{M_{\pi}^2}{f^2} F(z)\,,\\
F(z) &=& \int_1^{\infty} dx\, \frac{(x^2-1)^{3/2}}{x^2}  \Big( 2 B(x) +C(x)  - zx C(x)\Big)  e^{-z x}\,,
\end{eqnarray}
with 
\begin{eqnarray}
B(x) & = & \frac{8}{9} \left(g^2 + (g\tilde{u}  + v) x \right) \,,\\
C(x) & = & -\frac{8}{9} \left(g + \tilde{u}x \right)^2\,,
\end{eqnarray}
given in \pref{resultNLOBandBt} and \pref{resultNLOCcomb} but again expressed in terms of dimensionless quantities. In analogy to \pref{Defzandu} we have introduced
\begin{equation}
v\equiv \gamma M_{\pi}
\end{equation}
for the additional LEC that enters the NLO result. Expanding $C(x)$ and collecting powers of $x$ the function $F(z)$ can be expressed as the following sum involving the basic integrals $J_k(z)$:
\begin{eqnarray}
F(z)&=& \frac{8}{9}\Big( g^2J_0(z) +\big(g^2z+2v\big) J_1(z) +\big(2g\tilde{u}z - \tilde{u}^2\big) J_2(z) +\tilde{u}^2 zJ_3(z)\Big)\,.
\end{eqnarray}
\subsection{The master integrals $\mathbf{J_k(z)}$}
The integrals $J_k$ play the role of master integrals for the $B\pi$ contamination in the various observables considered in this paper. According to the previous section we need explicit expressions for $k=0,\ldots,3$.

As already mentioned, the $J_k$ can be expressed in terms of special mathematical functions. Explicitly, for the lowest four $k$ we find the following results:
\begin{eqnarray}
J_0(z) &=& \frac{3}{z} K_1(z) - z J_1(z)\,,\label{AppIntJ0}\\
J_1(z)&=& \frac{\pi}{2}\bigg( 1- z \left[ K_{2}(z) \mathbf{L}_{-3}(z) + K_{3}(z) \mathbf{L}_{-2}(z)\right]\bigg)\,,\label{AppIntJ1}\\
J_2(z) &=& \frac{3}{z^2}K_2(z)\,,\label{AppIntJ2}\\
J_3(z) &=& \frac{3}{z^2} K_3(z)\,.\label{AppIntJ3}
\end{eqnarray}
Here $K_n$ and $\mathbf{L}_{-n}$ denote the modified Bessel and Struve functions, respectively. For their precise definitions, properties, asymptotic expansions etc.\ the reader is referred to Refs.\ \cite{NIST:Handbook,NIST:DLMF}, chapters 10 and 11.

In order to establish that \pref{AppIntJ0} - \pref{AppIntJ3} are indeed the integrals defined in \pref{defJk} 
it is convenient to start with $k=2$. A standard integral representation for the modified Bessel function $K_{\nu}(z)$ reads \cite[\href{https://dlmf.nist.gov/10.32.E8}{(10.32.E8)}]{NIST:DLMF}
\begin{eqnarray}\label{defmodBesselK}
K_{\nu}(z) &=&\frac{\pi^{\frac{1}{2}}(\frac{1}{2}z)^{\nu}}{\Gamma\left(\nu+\frac{1}{2}\right)}\int_{1}^{\infty}e^{-zt}(t^{2}-1)^{\nu-\frac{1}{2}}\,\mathrm{d}t.
\end{eqnarray}
This representation holds for real $\nu> -1/2$ and for real $z$. Setting $\nu=2$ we easily find the result \pref{AppIntJ2}.

Two simple properties of $J_k(z)$ follow immediatly from the definition in eq.\ \pref{defJk}: 
\begin{eqnarray}
a) & & J_{k}(z) \,=\, -\frac{d}{dz} J_{k-1}(z)\,,\\[0.4ex]
b) & & \lim_{z\rightarrow\infty} J_k(z) \,=\, 0\,.
\end{eqnarray}
Together with known results \cite[\href{https://dlmf.nist.gov/10.29.E4}{(10.29.E4)}]{NIST:DLMF}), \cite[\href{https://dlmf.nist.gov/10.43.E2}{(10.43.E2)}]{NIST:DLMF} for derivatives and integrals of Besselfunctions it is straightforward to establish \pref{AppIntJ1} and \pref{AppIntJ3}. Finally, the right hand side of \pref{AppIntJ0} is the result of a simple partial integration of the integral \pref{defJk} with $k=0$.

\end{appendix}



\end{document}